\documentclass[reprint,superscriptaddress,prl]{revtex4-1}
\usepackage{amsmath}
\usepackage{amssymb}
\usepackage{bm}
\usepackage{braket}
\usepackage{color}
\usepackage[varg]{txfonts}
\usepackage{varwidth}
\usepackage{dcolumn}
\usepackage[breaklinks,colorlinks=true,linkcolor=blue,urlcolor=cyan,citecolor=blue]{hyperref}
\usepackage{graphicx}

\begin{document}

\title{Incommensurate broken helix induced by nonstoichiometry in the axion insulator candidate EuIn$_{2}$As$_{2}$}

\author{Masaki~Gen}
\email{gen@issp.u-tokyo.ac.jp}
\affiliation{RIKEN Center for Emergent Matter Science (CEMS), Wako 351-0198, Japan}
\affiliation{Institute for Solid State Physics, University of Tokyo, Kashiwa, Chiba 277-8581, Japan}

\author{Yukako~Fujishiro}
\affiliation{RIKEN Center for Emergent Matter Science (CEMS), Wako 351-0198, Japan}

\author{Kazuki~Okigami}
\affiliation{Department of Applied Physics, University of Tokyo, Tokyo 113-8656, Japan}

\author{Satoru~Hayami}
\affiliation{Graduate School of Science, Hokkaido University, Sapporo 060-0810, Japan}

\author{Max~T.~Birch}
\affiliation{RIKEN Center for Emergent Matter Science (CEMS), Wako 351-0198, Japan}

\author{Kiyohiro~Adachi}
\affiliation{Materials Characterization Support Team, RIKEN Center for Emergent Matter Science (CEMS), Wako 351-0198, Japan}

\author{Daisuke~Hashizume}
\affiliation{Materials Characterization Support Team, RIKEN Center for Emergent Matter Science (CEMS), Wako 351-0198, Japan}

\author{Takashi~Kurumaji}
\affiliation{Department of Advanced Materials Science, University of Tokyo, Kashiwa 277-8561, Japan}
\affiliation{Division of Physics, Mathematics and Astronomy, California Institute of Technology, Pasadena, California 91125, USA}

\author{Hajime~Sagayama}
\affiliation{Institute of Materials Structure Science, High Energy Accelerator Research Organization, Tsukuba 305-0801, Japan}

\author{Hironori~Nakao}
\affiliation{Institute of Materials Structure Science, High Energy Accelerator Research Organization, Tsukuba 305-0801, Japan}

\author{Yoshinori~Tokura}
\affiliation{RIKEN Center for Emergent Matter Science (CEMS), Wako 351-0198, Japan}
\affiliation{Department of Applied Physics, University of Tokyo, Tokyo 113-8656, Japan}
\affiliation{Tokyo College, University of Tokyo, Tokyo 113-8656, Japan}

\author{Taka-hisa~Arima}
\affiliation{RIKEN Center for Emergent Matter Science (CEMS), Wako 351-0198, Japan}
\affiliation{Department of Advanced Materials Science, University of Tokyo, Kashiwa 277-8561, Japan}

\begin{abstract}

Zintl phase EuIn$_{2}$As$_{2}$ has garnered growing attention as an axion insulator candidate, triggered by the identification of a commensurate double-${\mathbf Q}$ broken-helix state in previous studies, however, its periodicity and symmetry remain subjects of debate.
Here, we perform resonant x-ray scattering experiments on EuIn$_{2}$As$_{2}$, revealing an incommensurate nature of the broken-helix state, where both the wave number and the amplitude of the helical modulation exhibit systematic sample dependence.
Furthermore, the application of an in-plane magnetic field brings about a fanlike state that appears to preserve the double-${\mathbf Q}$ nature, which might be attributed to multiple-spin interactions in momentum space.
We propose that the itinerant character of EuIn$_{2}$As$_{2}$, most likely induced by Eu deficiency, gives rise to the helical modulation and impedes the realization of a theoretically-predicted axion state with the collinear antiferromagnetic order.

\end{abstract}

\date{\today}
\maketitle

Europium-based compounds offer a fertile playground for exploring nontrivial magnetotransport phenomena \cite{2004_Wig, 2016_Kaw, 2020_Ros, 2023_May}.
In the presence of Fermi surfaces, the carrier-mediated Ruderman-Kittel-Kasuya-Yosida (RKKY) interaction can stabilize a variety of modulated magnetic structures such as helix \cite{2009_Nan, 2020_Tak, 2022_Kur, 2023_Soh_2} and skyrmions \cite{2019_Kan, 2022_Tak, 2023_Gen, 2023_Sin, 2023_Mat}.
The correlation between magnetism and electronic band topology is one other intriguing aspect, as extensively studied in the 122 families of the Zintl phase \cite{2006_Jia, 2019_Ma, 2020_Su, 2021_Wan, 2022_Bia, 2023_Kre, 2023_Cuo}.
First-principles calculations predict that EuIn$_{2}$As$_{2}$ can host an axion insulating state (AXS) with the layered antiferromagnetic (AFM) order \cite{2009_Ess, 2019_Xu}, thereby potentially exhibiting the quantized magnetoelectric effect \cite{2019_Tok, 2022_Ber, 2021_Fij}.
Given that the AXS remains unestablished in stoichiometric bulk materials \cite{2021_Fij, 2011_Wan, 2017_Mog, 2019_Zha, 2019_Li, 2020_Den}, a comprehensive  investigation on the physical properties of EuIn$_{2}$As$_{2}$ stands as a critical issue \cite{2008_Gof, 2020_Zha, 2020_Sat, 2020_Reg, 2020_Yu, 2021_Rib, 2022_Gon, 2022_Yan, 2023_Soh, 2023_Wu, 2023_Don}, not only from the fundamental viewpoint but also for its applications in next-generation devices \cite{2019_Tok}.

\begin{figure}[t]
\centering
\includegraphics[width=0.95\linewidth]{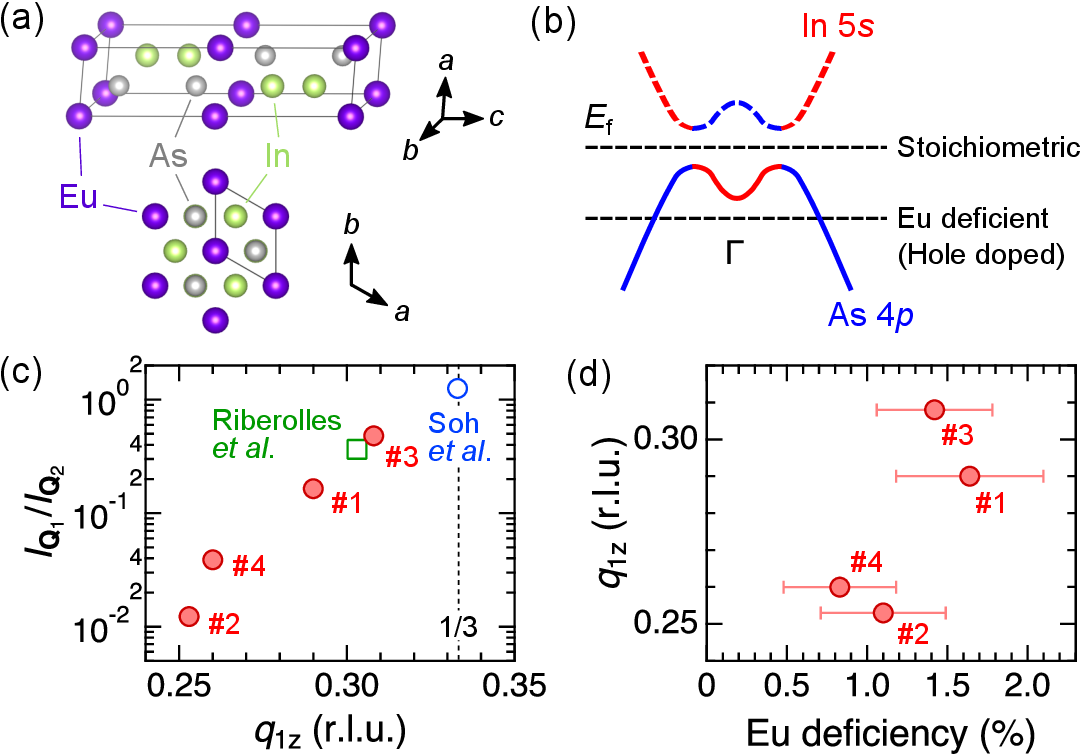}
\caption{(a) Crystal structure of Zintl phase EuIn$_{2}$As$_{2}$. The black line represents a crystallographic unit cell, which contains two Eu layers. (b) Schematic of the bulk band structure around the $\Gamma$ point. The Fermi level is in the gap according to the first-principles calculations \cite{2019_Xu, 2023_Cao}, although a hole pocket was observed in ARPESs \cite{2020_Sat, 2020_Reg}. (c) Relationship between the $q_{1z}$ value and the integrated-intensity ratio of the ${\mathbf Q}_{1}$ to the ${\mathbf Q}_{2}$ peak, $I_{{\mathbf Q}_{1}}/I_
{{\mathbf Q}_{2}}$, observed for samples \#1 $\sim$ \#4 in this study (5~K) and those reported in Refs.~\cite{2021_Rib, 2023_Soh} (6~K). See the SM \cite{SM} for the correction factors in estimating $I_{{\mathbf Q}_{1}}/I_{{\mathbf Q}_{2}}$. (d) Plots of the $q_{1z}$ value versus Eu deficiency for samples \#1 $\sim$ \#4.}
\vspace{-0.2cm}
\label{Fig1}
\end{figure}

EuIn$_{2}$As$_{2}$ forms a hexagonal crystal structure with the space group $P6_{3}/mmc$, consisting of alternating stacks of Eu triangular layers and In$_{2}$As$_{2}$ blocks [Fig.~\ref{Fig1}(a)].
The theoretically-predicted band structure is schematically illustrated in Fig.~\ref{Fig1}(b) \cite{2019_Xu, 2023_Cao}.
There is a crossing of In $5s$ and As $4p$ bands near the $\Gamma$ point, where a gap opens due to spin-orbit coupling.
Although the Fermi energy is theoretically located in the band gap, angle-resolved photoemission spectroscopies (ARPESs) revealed holelike bulk bands crossing the Fermi level \cite{2020_Sat, 2020_Reg}.
The magnetism is dominated by intraplane ferromagnetic and interplane AFM interactions between localized Eu$^{2+}$ moments, with an easy-plane anisotropy becoming evident at low temperatures \cite{2020_Zha}. 
Neutron diffraction (ND) \cite{2021_Rib} and resonant x-ray scattering (RXS) experiments \cite{2023_Soh} observed a concurrent short-period magnetic modulation, ${\mathbf Q}_{1} = (0, 0, q_{1z})$, along with the AFM component ${\mathbf Q}_{2} = (0, 0, 1)$; the reported $q_{1z}$ values are 0.303(1) \cite{2021_Rib} and 0.3328(6) \cite{2023_Soh}, respectively, in the reciprocal-lattice unit.
A phase separation scenario was excluded by azimuthal scans in the RXS \cite{2023_Soh}.
Both studies proposed a six-layer-period double-${\mathbf Q}$ helical structure, termed {\it broken helix}, under the assumption that ${\mathbf Q}_{1}$ represents an exactly commensurate modulation with $q_{1z} = 1/3$ \cite{2021_Rib, 2023_Soh}.
Initially, the AXS was believed to be realized in the commensurate broken-helix state, as the ${\mathcal T}{\mathcal C}_{2}$ symmetry (the combination of time-reversal and two-fold rotational operations) is preserved once the principal axis of the broken helix is aligned along a specific crystallographic axis \cite{2021_Rib, 2023_Soh}.
However, a recent optical birefringence study challenged the above picture in terms of the symmetry, and proposed the unpinned nature of the broken helix owing to minimal hexagonal anisotropy \cite{2023_Don}.

In this Letter, we reexamine the magnetic structure of EuIn$_{2}$As$_{2}$.
While the preceding ND study identified an {\it incommensurate} ${\mathbf Q}_{1}$ modulation \cite{2021_Rib}, no incommensurate spin configuration has been considered so far \cite{2021_Rib, 2023_Soh, 2023_Don}.
Whether ${\mathbf Q}_{1}$ is commensurate or not is pivotal, as the latter is not compatible with the AXS due to the ${\mathcal T}{\mathcal C}_{2}$ symmetry breaking.
The mechanism for the emergence of the ${\mathbf Q}_{1}$ modulation also remains puzzling \cite{2023_Soh, 2023_Don}.
To address these issues, we perform the RXS experiments, revealing that the ${\mathbf Q}_{1}$ peak exhibits variability in the $q_{1z}$ value (\textcolor{blue}{0.25--0.31}) as well as the intensity across samples [Fig.~\ref{Fig1}(c)].
The complementary single-crystal structure analyses reveal a greater amount of Eu deficiency for samples exhibiting a larger $q_{1z}$ value [Fig.~\ref{Fig1}(d)].
These results suggest that the ${\mathbf Q}_{1}$ modulation stems from the RKKY interaction mediated by doped holes due to Eu deficiency.
We argue that, to avoid the loss of generality, the broken-helix state should be considered as a superposition of incommensurate helical and collinear AFM modulations.
The appearance of an exotic double-${\mathbf Q}$ fanlike state is also revealed  in an in-plane magnetic field.

\begin{figure}[t]
\centering
\includegraphics[width=0.95\linewidth]{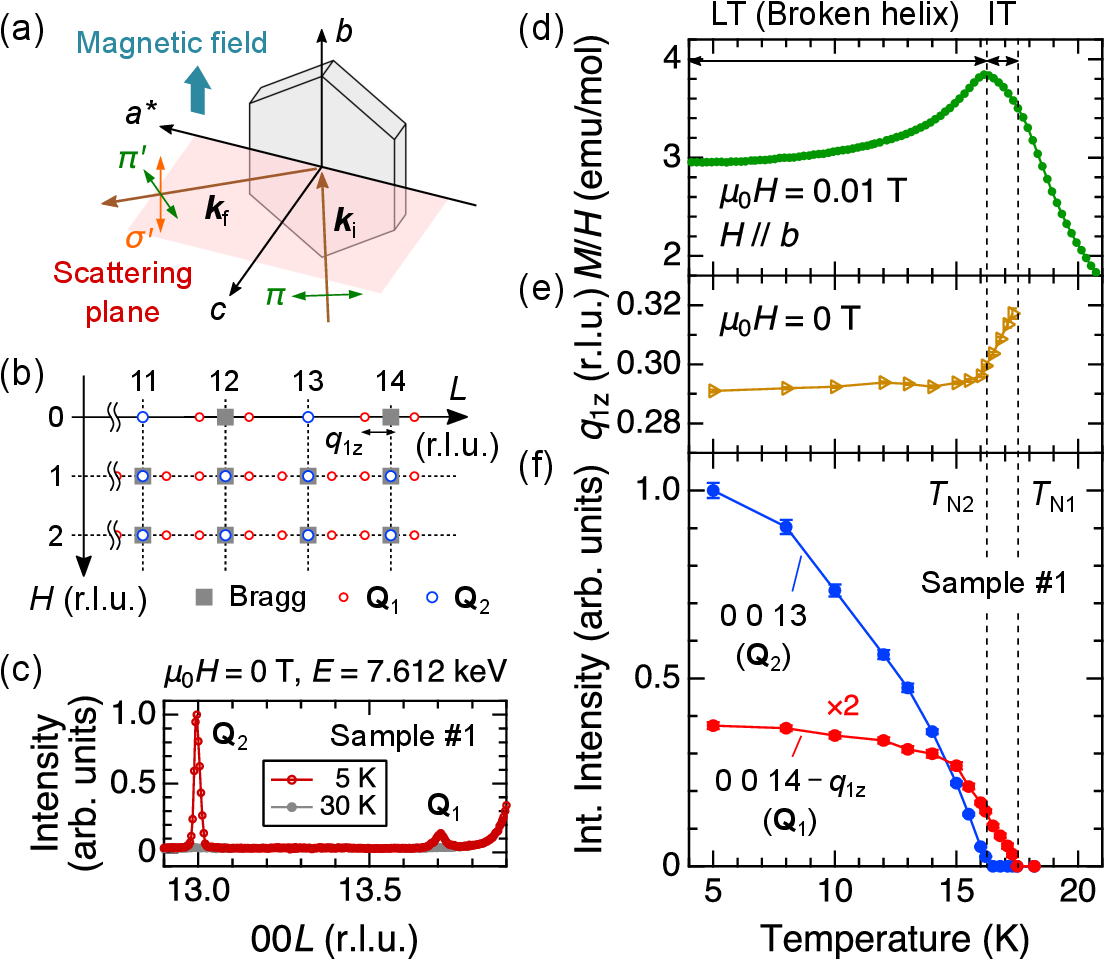}
\caption{(a) Experimental geometry of the RXS. ${\mathbf k}_{\rm i}$ (${\mathbf k}_{\rm f}$) and $\pi$ ($\pi'$ and $\sigma'$) represent the propagation vector and polarization direction of incident (scattered) x-rays, respectively. (b) Positions of the Bragg peaks in the $(H, 0, L)$ scattering plane below $T_{\rm N2}$. Gray squares are the fundamental peaks, and red (blue) open circles are the magnetic ${\mathbf Q}_{1}$ (${\mathbf Q}_{2}$) peaks. (c) RXS profiles observed in the $(0, 0, L)$ scan at 5~K (red) and 30~K (gray) at zero field for sample \#1. [(d)--(f)] Temperature dependence of (d) magnetic susceptibility $M/H$, (e) $q_{1z}$, and (f) integrated intensity of the $00L$ reflections with $L = 14 - q_{1z}$ (red) and $L = 13$ (blue) at zero field.}
\vspace{-0.2cm}
\label{Fig2}
\end{figure}

Single crystals of EuIn$_{2}$As$_{2}$ were grown by an indium flux method.
Details of sample growth and characterization are described in the Supplemental Material (SM) \cite{SM}.
We picked up four specimens (\#1 $\sim$ \#4) from the same batch.
RXS experiments were performed at BL-3A, Photon Factory, KEK, by using horizontally ($\pi$) polarized incident x-rays in resonance with the Eu $L_{2}$ absorption edge ($E = 7.612$~keV) \cite{2022_Kur, 2023_Gen}.
The incident x-ray beam was collimated with $1 \times 1$~mm$^{2}$, which is comparable or slightly larger than the sample size.
Accordingly, the observed RXS intensity would come from almost all area of the sample.
Samples with the as-grown (001) plane were glued on an aluminum plate and set in a cryostat equipped with a vertical-field superconducting magnet.
The scattering plane was set to be $(H, 0, L)$, and a magnetic field was applied along the crystallographic $b$ axis [Fig.~\ref{Fig2}(a)].
We could access fundamental Bragg peaks at $(3m, 0, 2n)$ and $(3m \pm 1, 0, n)$ ($m, n$: integer), and their magnetic satellite peaks in the reciprocal space [Fig.~\ref{Fig2}(b)].
Unless otherwise stated, we performed the $(0, 0, L)$ scan with $L=12$--16, and the scattered x-rays were detected without analyzing the $\pi'$ and $\sigma'$ polarizations, parallel and perpendicular to the scattering plane, respectively.
For polarization analysis, the 006 reflection of a pyrolytic graphite (PG) crystal was used, where the scattering angle was $\sim$92$^{\circ}$ near the Eu $L_{2}$ edge.

As in Refs.~\cite{2021_Rib, 2023_Soh}, we observe two kinds of magnetic Bragg peaks, ${\mathbf Q}_{1} = (0, 0, q_{1z})$ and ${\mathbf Q}_{2} = (0, 0, 1)$ [Fig.~\ref{Fig2}(c)].
Figures~\ref{Fig2}(d)--\ref{Fig2}(f) show the temperature dependence of magnetic susceptibility, $q_{1z}$, and integrated intensities of the ${\mathbf Q}_{1}$ and ${\mathbf Q}_{2}$ peaks at (nearly) zero field for sample \#1.
Upon cooling, the ${\mathbf Q}_{1}$ and ${\mathbf Q}_{2}$ peaks emerge below $T_{\rm N1} = 17.6$~K and $T_{\rm N2} = 16.2$~K, respectively.
$q_{1z}$ gradually decreases from 0.32 to 0.30 in the intermediate-temperature (IT) phase, while $q_{1z}$ is almost temperature independent in the low-temperature (LT) phase, eventually reaching $q_{1z} = 0.29$ at 5~K.
These features agree with Refs.~\cite{2021_Rib, 2023_Soh}, although there is a discrepancy in the $q_{1z}$ value.
Notably, distinct $q_{1z}$ values are found for the other three samples: $q_{1z} = 0.25, 0.31$ and 0.26 for samples \#2, \#3, and \#4, respectively (for additional RXS data, see the SM \cite{SM}).
Besides, we find an important trend that the integrated intensity ratio of the ${\mathbf Q}_{1}$ peak to the ${\mathbf Q}_{2}$ peak, $I_{{\mathbf Q}_{1}}/I_
{{\mathbf Q}_{2}}$, is higher for samples with larger $q_{1z}$ [Fig.~\ref{Fig1}(c)].
We note that the way of adhering samples to the Al plate had little influence on the RXS profile.
Therefore, the observed variability in $q_{1z}$ and $I_{{\mathbf Q}_{1}}/I_
{{\mathbf Q}_{2}}$ appears to be primarily attributed to the sample dependence rather than extrinsic strain.

To confirm this from the structural point of view, we performed single-crystal x-ray diffraction experiments for all the four samples.
Our crystal-structure analyses reveal the presence of a Eu deficiency of 1.64(46)\%, 1.10(39)\%, 1.42(36)\%, and 0.83(35)\% for samples \#1 $\sim$ \#4, respectively (for details, see the SM \cite{SM}).
As shown in Fig.~\ref{Fig1}(d), one can see a trend that samples exhibiting a larger $q_{1z}$ value host a greater amount of Eu deficiency.
Recalling the first-principles calculations predicting a gapped insulating state with the AFM order characterized by ${\mathbf Q}_{2}$ \cite{2019_Xu}, the additional ${\mathbf Q}_{1}$ modulation would be induced by doped hole carriers which contribute to the RKKY interaction between the Eu$^{2+}$ moments.
The above scenarios agree with the three-dimensional character of a hole pocket observed in a previous ARPES \cite{2020_Sat}, although the details of the Fermi-surface shape need to be clarified.
Indeed, our Hall resistivity measurements reveal the multi-carrier nature, making the quantitative estimation of the carrier number challenging (see the SM \cite{SM}).

\begin{figure}[t]
\centering
\includegraphics[width=0.95\linewidth]{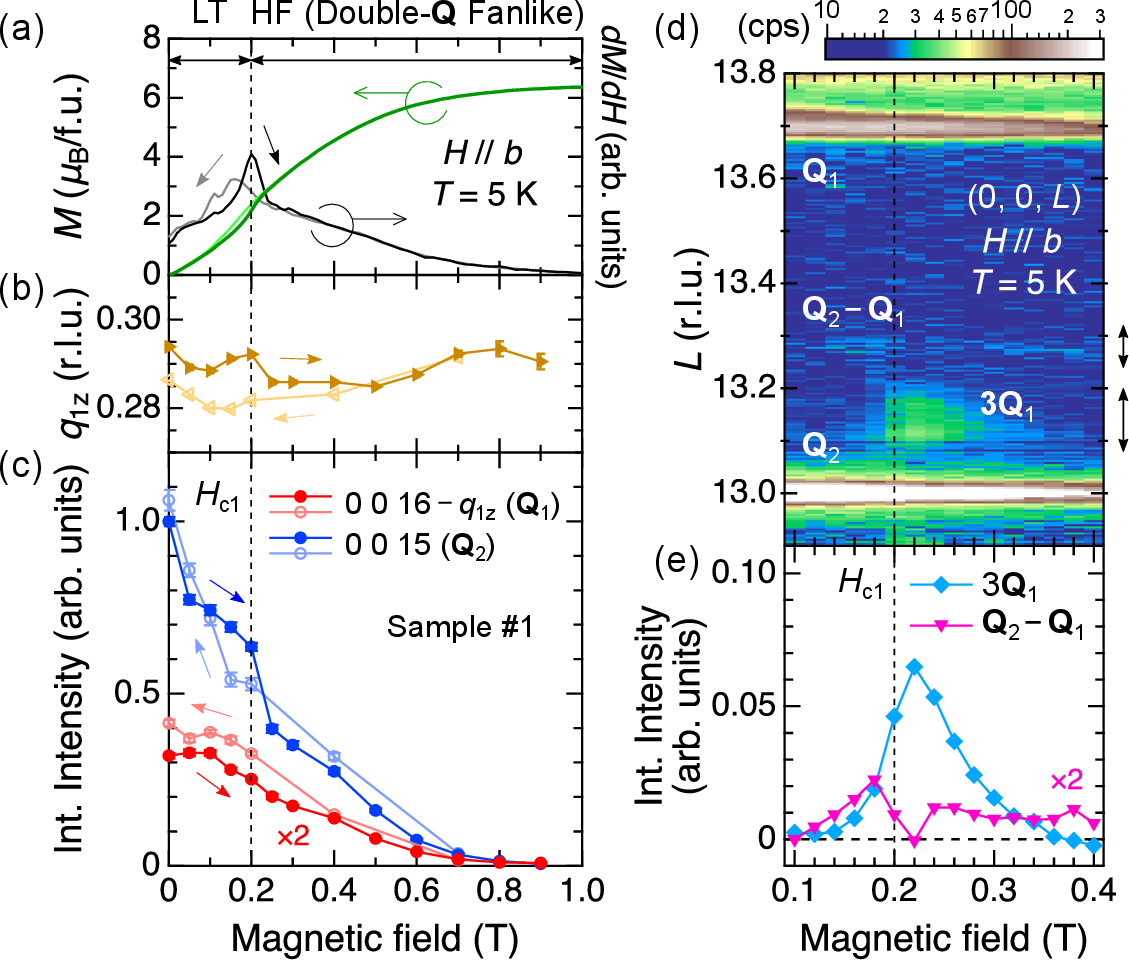}
\caption{[(a)--(c)] Magnetic-field dependence of (a) magnetization $M$ (left axis) and its field derivative $dM/dH$ (right axis), (b) $q_{1z}$, and (c) integrated intensity of the $00L$ reflections with $L = 16-q_{1z}$ (red) and $L = 15$ (blue). Dark and light colors represent data in field-increasing and decreasing processes, respectively. (d) Logarithmic contour plot of the RXS intensity in the $(0, 0, L)$ scan near $H_{\rm c1}$ in a field-increasing process. Black arrows on the right represent peak widths for $3{\mathbf Q}_{1}$ and ${\mathbf Q}_{2}-{\mathbf Q}_{1}$. (e) Magnetic-field dependence of the integrated intensity of the $3{\mathbf Q}_{1}$ (cyan) and ${\mathbf Q}_{2}-{\mathbf Q}_{1}$ (pink) peaks. The absolute value is normalized by the intensity of the 0013 reflection at 0~T. All the data are taken at 5~K with $H \parallel b$ for sample \#1.}
\vspace{-0.2cm}
\label{Fig3}
\end{figure}

Next, we investigate the magnetic structure changes with the application of an in-plane magnetic field for sample \#1 ($q_{1z} = 0.29$), which undergoes a metamagnetic transition at $\mu_{0}H_{\rm c1} = 0.2$~T at 5~K \cite{2020_Zha, 2020_Yu, 2023_Soh}.
Figures~\ref{Fig3}(a)--(c) show the field dependence of magnetization, $q_{1z}$, and integrated intensities of the ${\mathbf Q}_{1}$ and ${\mathbf Q}_{2}$ peaks.
With increasing a magnetic field, both ${\mathbf Q}_{1}$ and ${\mathbf Q}_{2}$ peaks survive until the saturation field of 1.0~T.
Here, $q_{1z}$ exhibits only a small change of at most $\pm 0.01$ even across $H_{\rm c1}$.
With subsequently decreasing a magnetic field, a hysteresis appears around $H_{\rm c1}$ in the RXS profile, in line with the magnetization process.
Ultimately, neither the intensities of the ${\mathbf Q}_{1}$ and ${\mathbf Q}_{2}$ peaks nor $q_{1z}$ revert to the original values at zero field.
The observed intensity changes against a magnetic field are qualitatively consistent with Ref.~\cite{2023_Soh}.

We find remarkable features in the RXS profile near $H_{\rm c1}$.
Figure~\ref{Fig3}(d) shows a logarithmic contour plot of the RXS intensity against a magnetic field, observed in the $(0, 0, L)$ scan in the $L$ range between 12.9 and 13.8.
Apart from the strong ${\mathbf Q}_{1}$ and ${\mathbf Q}_{2}$ peaks at $L=13.71$ and 13, respectively, peaks around $L=13.29$ and 13.13 are discernible in certain field ranges (for the RXS profile, see the SM \cite{SM}).
These additional peaks likely correspond to higher-harmonic ${\mathbf Q}_{2}-{\mathbf Q}_{1}$ and $3{\mathbf Q}_{1}$ modulations, respectively.
The field evolution of integrated intensities of these peaks are displayed in Fig.~\ref{Fig3}(e).
The weak ${\mathbf Q}_{2}-{\mathbf Q}_{1}$ peak appears above 0.1~T and persists until at least 0.4~T, except immediately after $H_{\rm c1}$ where the intensity of the $3{\mathbf Q}_{1}$ peak becomes prominent.
The presence of the ${\mathbf Q}_{2}-{\mathbf Q}_{1}$ peak even above $H_{\rm c1}$ indicates the double-${\mathbf Q}$ nature with a superposition of ${\mathbf Q}_{1}$ and ${\mathbf Q}_{2}$ in the high-field (HF) phase.
We note that the $2{\mathbf Q}_{1}$ peak, expected in the conventional fanlike structure \cite{2003_Kos, 2020_Ghi}, is not observed at $L = 13.42$ in the entire field range, in spite of the observation of the $3{\mathbf Q}_{1}$ peak [Fig.~\ref{Fig3}(d)].
This also rules out the possibility of the presence of a single-${\mathbf Q}$ fanlike state in the HF phase.

\begin{figure}[t]
\centering
\includegraphics[width=0.95\linewidth]{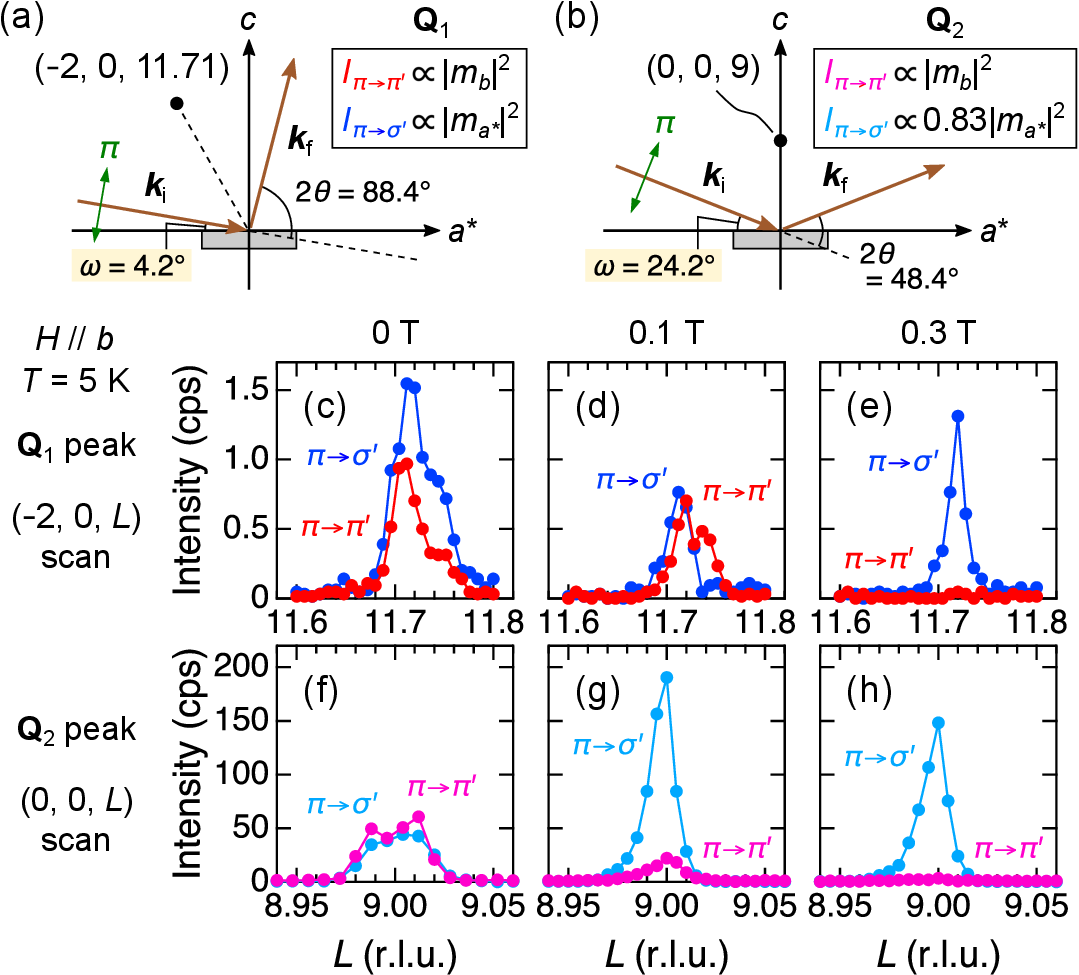}
\caption{[(a)(b)] Schematic geometical configuration of the RXS focusing on the ${\mathbf Q}_{1}$ and ${\mathbf Q}_{2}$ peaks around the magnetic Bragg spots $(-2, 0, 11.71)$ and $(0, 0, 9)$, respectively. See also Figs.~\ref{Fig2}(a) and \ref{Fig2}(b). [(c)--(h)] RXS profiles of the polarization analysis at 5~K at 0~T [(c)(f)], 0.1~T [(d)(g)], and 0.3~T [(e)(h)] with $H \parallel b$. Panels (c)--(e) and (f)--(h) show the data of the $(-2, 0, L)$ scan around $L = 11.71$ and the $(0, 0, L)$ scan around $L = 9$, respectively.}
\vspace{-0.2cm}
\label{Fig4}
\end{figure}

To get information on the orientation of magnetic moments in each phase at 5~K, we performed polarization analysis with $H \parallel b$.
In general, the magnetic scattering intensity $I$ is given by $I \propto |({\mathbf e}_{\rm i} \times {\mathbf e}_{\rm f}) \cdot {\mathbf m}_{\mathbf Q}|^{2}$, where ${\mathbf e}_{\rm i}$ (${\mathbf e}_{\rm f}$) is the polarization vector of the incident (scattered) beam, and ${\mathbf m}_{\mathbf Q}$ is a magnetic moment of the ${\mathbf Q}$ modulation.
The $\pi$--$\pi'$ channel ($I_{\pi-\pi'}$) always detects the modulated component along $b$ ($m_{b}$), whereas the $\pi$--$\sigma'$ channel ($I_{\pi-\sigma'}$) contains those along $a^{*}$ ($m_{a^{*}}$) and $c$ ($m_{c}$) in the ratio of $\cos^{2}\omega : \sin^{2}\omega$, where $\omega$ is the angle between the propagation vector of the incident beam (${\mathbf k}_{\rm i}$) and the $a^{*}$ axis.
Figures~\ref{Fig4}(a) and \ref{Fig4}(b) show the experimental configurations of polarization analysis focusing on the Bragg spots at $(-2, 0, 11.71)$ and $(0, 0, 9)$, in which $I_{\pi-\sigma'}$ mainly reflects $m_{a^{*}}$ ($\sim$100\% and 83\%) of the ${\mathbf Q}_{1}$ and ${\mathbf Q}_{2}$ modulations, respectively.
The results are summarized in Figs.~\ref{Fig4}(c)--\ref{Fig4}(h).
Additional measurements focusing on the Bragg spots at $(2, 0, 11.71)$ and $(0, 0, 17)$ confirm the absence of $m_{c}$ (see the SM \cite{SM}), i.e., all the spins lie within the $ab$ plane.

At zero field, both $I_{\pi-\pi'}$ and $I_{\pi-\sigma'}$ have intensities at $(-2, 0, 11.71)$ and $(0, 0, 9)$ [Figs.~\ref{Fig4}(c) and \ref{Fig4}(f)].
At 0.1~T ($< H_{\rm c1}$), a drastic enhancement is observed in $I_{\pi-\sigma'}$ at $(0, 0, 9)$, accompanied by a suppression of $I_{\pi-\pi'}$ [Fig.~\ref{Fig4}(g)], suggesting the reorientation of magnetic domains so as to lie the AFM component perpendicular to the field direction.
In contrast, $I_{\pi-\pi'}$ and $I_{\pi-\sigma'}$ still have comparable intensities at $(-2, 0, 11.71)$ [Fig.~\ref{Fig4}(d)].
This observation suggests that the ${\mathbf Q}_{1}$ peak originates from a helical modulation rather than a sinusoidal one, as the intensity should appear only in either $I_{\pi-\pi'}$ or $I_{\pi-\sigma'}$ in the latter case.
Followed by the metamagnetic transition, the presence of only $m_{a^{*}}$ is confirmed for both the ${\mathbf Q}_{1}$ and ${\mathbf Q}_{2}$ modulations, as evidenced by the disappearance of $I_{\pi-\pi'}$ at both Bragg positions at 0.3~T ($> H_{\rm c1}$) [Figs.~\ref{Fig4}(e) and \ref{Fig4}(h)].
Accordingly, the HF phase can be ascribed to a double-${\mathbf Q}$ fanlike state.

\begin{figure}[t]
\centering
\includegraphics[width=0.95\linewidth]{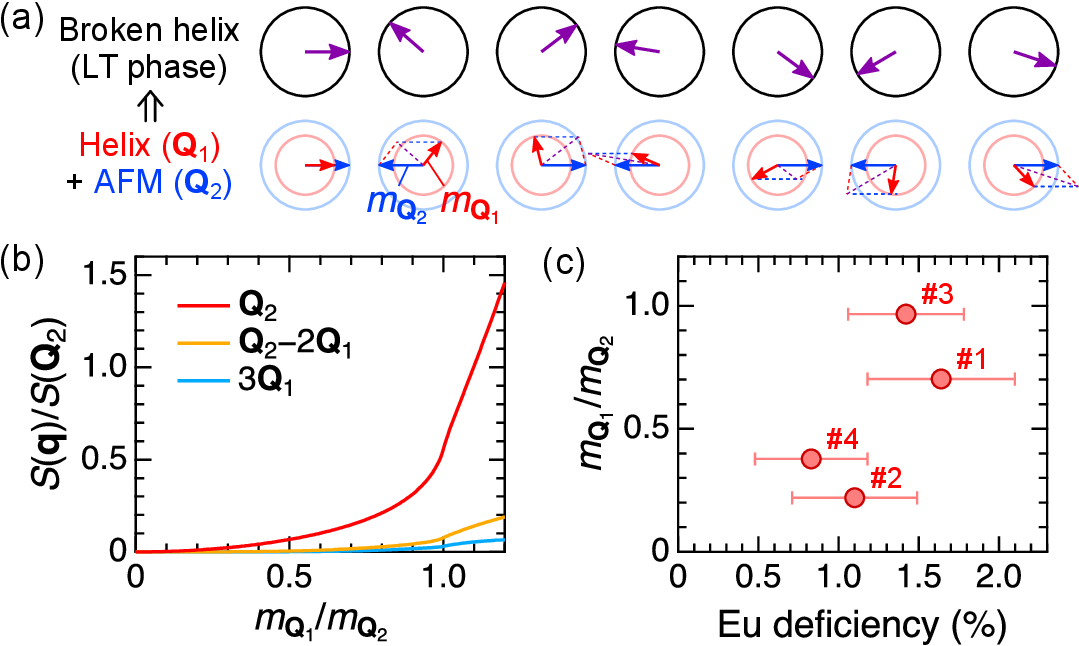}
\caption{(a) Schematic of the broken-helix state and an equivalent double-${\mathbf Q}$ representation as the superposition of incommensurate helical ${\mathbf Q}_{1}$ and collinear AFM ${\mathbf Q}_{2}$ modulations. Magnetic moments are in the $ab$-plane, and the modulation is along the $c$ axis. (b) $m_{{\mathbf Q}_{1}}/m_{{\mathbf Q}_{2}}$ dependence of the spin structure factor $S({\mathbf q})$ relative to $S({\mathbf Q}_{2})$ at ${\mathbf Q}_{1}$, ${\mathbf Q}_{2} - 2{\mathbf Q}_{1}$, and $3{\mathbf Q}_{1}$. The $S({\mathbf q})$ profiles exhibit an anomaly at a singular point $m_{{\mathbf Q}_{1}}/m_{{\mathbf Q}_{2}} = 1$, where the antiparallel $m_{{\mathbf Q}_{1}}$ and $m_{{\mathbf Q}_{2}}$ moments cancel each other out. (c) Relationship between Eu deficiency and a moment-amplitude ratio of the ${\mathbf Q}_{1} = (0, 0, q_{1z})$ to the ${\mathbf Q}_{2} = (0, 0, 1)$ modulation, $m_{{\mathbf Q}_{1}}/m_{{\mathbf Q}_{2}}$, in the broken-helix state for samples \#1 $\sim$ \#4.}
\vspace{-0.2cm}
\label{Fig5}
\end{figure}

Let us here update the understanding of the magnetic structure in EuIn$_{2}$As$_{2}$.
Our RXS experiments suggest that the LT phase is a broken-helix state with a superposition of incommensurate helical ${\mathbf Q}_{1}$ and AFM ${\mathbf Q}_{2}$ modulation [Fig.~\ref{Fig5}(a)], which can be approximately described as ${\mathbf S}_{i}$ = ${\mathbf m}({\mathbf r}_{i})/|{\mathbf m}({\mathbf r}_{i})|$, where ${\mathbf m}({\mathbf r}_{i}) \propto (1, i, 0) \sum_{\eta = 1, 2}{m_{{\mathbf Q}_{\eta}}}\exp(i{\mathbf Q}_{\eta}\cdot{\mathbf r}_{i}) + {\rm c.c.}$ at zero field.
This expression encompasses the commensurate broken helix proposed in Refs.~\cite{2021_Rib, 2023_Soh} as a special case.
Based on the above formula, we calculate the spin structure factor, $S({\mathbf q}) = (1/N)\sum_{i,j}\langle {\mathbf S}_{i} \cdot {\mathbf S}_{j}\rangle e^{i{\mathbf q}({\mathbf r}_{i}-{\mathbf r}_{j})}$, at ${\mathbf Q}_{1}$ and ${\mathbf Q}_{2}$ as a function of $m_{{\mathbf Q}_{1}}/m_{{\mathbf Q}_{2}}$, as shown in Fig.~\ref{Fig5}(b) \cite{comment}.
By comparing the calculated $S({\mathbf Q}_{1})/S({\mathbf Q}_{2})$ with the observed intensity ratio $I_{{\mathbf Q}_{1}}/I_
{{\mathbf Q}_{2}}$ [Fig.~\ref{Fig1}(c)], we estimate $m_{{\mathbf Q}_{1}}/m_{{\mathbf Q}_{2}}$ at the lowest temperature for each sample used in this study, as shown in Fig.~\ref{Fig5}(c) (for details, see the SM \cite{SM}).
A positive correlation between the Eu deficiency and $m_{{\mathbf Q}_{1}}/m_{{\mathbf Q}_{2}}$ agrees with the RKKY picture for the emergence of helimagnetism by hole doping, as mentioned above.
Furthermore, our simulation of $S({\mathbf q})$ for the broken-helix state indicates the appearance of weak higher-harmonic peaks at ${\mathbf Q}_{2}-2{\mathbf Q}_{1}$ and $3{\mathbf Q}_{1}$ [Fig.~\ref{Fig5}(b)], although we could not confirm their presence in the RXS experiments at zero field.
We note that the IT phase was identified as a sinusoidal state by M\"{o}ssbauer \cite{2021_Rib} and optical birefringence measurements \cite{2023_Don}.
As the hexagonal anisotropy is negligibly weak \cite{2020_Zha, 2023_Don}, the sinusoidal modulation would be stabilized by thermal fluctuations and gradually transforms into helix with decreasing temperature.
We thus infer that in the LT phase at 5~K, there is a slight elliptical distortion of the helix component, leading to the suppression of the higher harmonics.

Our RXS experiments also suggest a helix-fan transition while keeping the double-${\mathbf Q}$ nature in an in-plane magnetic field, as indicated by the observation of higher harmonics ${\mathbf Q}_{2} - {\mathbf Q}_{1}$ and $3{\mathbf Q}_{1}$, along with the absence of $2{\mathbf Q}_{1}$.
There have been reports of other Eu-based itinerant magnets exhibiting the coexistence of two magnetic modulations along the $c$ axis, such as EuRh$_{2}$As$_{2}$ \cite{2009_Nan}, EuCuSb \cite{2020_Tak}, and EuZnGe \cite{2022_Kur}, although their microscopic origins as well as field-induced magnetic structure changes are not yet fully understood.
Accordingly, it is crucial to establish theoretical approaches for a better understanding of the complex helimagnetism characterized by double-${\mathbf Q}$ interplane modulations.
Here, we investigate several types of spin models, aiming to reproduce the field-induced phase transition in EuIn$_{2}$As$_{2}$.
The detailed calculation results are presented in the SM \cite{SM}, and in the following, we briefly describe our updated knowledge.

Donoway {\it et al.} proposed a spin Hamiltonian composed of long-range Heisenberg exchange and biquadratic exchange interactions in real space on a one-dimensional chain \cite{2023_Don}.
Through simulated annealing, we find that this model can stabilize an incommensurate broken helix with a dominant ${\mathbf Q}_{2}$ component at zero field.
When applying an in-plane magnetic field, however, it fails to preserve the double-${\mathbf Q}$ nature and instead stabilizes a canted AFM state.
Alternatively, an effective spin Hamiltonian composed of bilinear and biquadratic exchange interactions in momentum space \cite{2017_Hay} can yield qualitative agreement with the experimental observations, provided that some additional terms are introduced: (i) an intertwined exchange coupling term, $-K_{2}({\mathbf S}_{{\mathbf Q}_{1}}\cdot{\mathbf S}_{{\mathbf Q}_{2}})({\mathbf S}_{-{\mathbf Q}_{1}}\cdot{\mathbf S}_{-{\mathbf Q}_{2}})$, and (ii) higher-harmonic exchange coupling terms, $-J''{\mathbf S}_{3{\mathbf Q}_{1}}\cdot{\mathbf S}_{-3{\mathbf Q}_{1}}$ and $K''({\mathbf S}_{3{\mathbf Q}_{1}}\cdot{\mathbf S}_{-3{\mathbf Q}_{1}})^{2}$, where $K_{2}, J'', K'' > 0$.
These terms (i) and (ii) are necessary to stabilize a double-${\mathbf Q}$ fanlike state and to enhance the $3{\mathbf Q}_{1}$ modulation, respectively, in high magnetic fields.
The resultant magnetic structure exhibits a peculiar feature of the double-${\mathbf Q}$ fanlike state.
Immediately after the metamagnetic transition, a square-wave-shaped modulation appears perpendicular to the field direction, reflecting the effect of $3{\mathbf Q}_{1}$ along with ${\mathbf Q}_{1}$.
In higher magnetic fields, where the $3{\mathbf Q}_{1}$ modulation diminishes, the spin configuration transforms into a conventional fanlike state, while the remaining ${\mathbf Q}_{2}$ component contributes to a complex spin-flipping pattern.

In summary, we elucidate the incommensurate nature of the double-${\mathbf Q}$ broken-helix state in EuIn$_{2}$As$_{2}$ through RXS experiments, contradicting the previously proposed commensurate broken-helix \cite{2021_Rib, 2023_Soh}.
Furthermore, we observe the possible emergence of a double-${\mathbf Q}$ fanlike state with higher-harmonic modulations in an in-plane magnetic field.
The double-${\mathbf Q}$ nature likely arises from the RKKY mechanism via hole carriers introduced by Eu deficiency.
To verify this scenario, further detailed studies on multiple samples, including ARPES, magnetotransport measurements, and extended x-ray absorption fine structure analysis, would be beneficial.
Our study indicates that the modulation period of the helix component is an excellent indicator for evaluating the sample quality of EuIn$_{2}$As$_{2}$.
We propose that electron doping in {\it off-stoichiometric} EuIn$_{2}$As$_{2}$ through chemical substitution can be a promising pathway to shift the Fermi energy to the band gap and suppress the helical modulation, realizing the theoretically predicted AXS with the collinear AFM order \cite{2019_Xu}.

This work was financially supported by the JSPS KAKENHI Grants-In-Aid for Scientific Research (No.~22H00101, No.~22K14011, No.~23H04869, No.~23H05431, and No.~23K13068), JST SPRING (No.~JPMJSP2108), and JST PRESTO (No.~MJPR20L8).
The resonant x-ray scattering experiments were performed with the approval of the Photon Factory Program Advisory Committee (Proposal No.~2022G551 and No.~2023G093).
The authors appreciate A. Kikkawa and M. Kriener for the invaluable support in the synthesis of the single crystals used in this study.
The authors appreciate H. Takeda, S. Kitou, M. Hirschberger, and K. Kakurai for fruitful discussions.


\begin{thebibliography}{99}
\bibitem{2004_Wig} G. A. Wigger, C. Beeli, E. Felder, H. R. Ott, A. D. Bianchi, and Z. Fisk, Percolation and the Colossal Magnetoresistance of Eu-Based Hexaboride, Phys. Rev. Lett. {\bf 93}, 147203 (2004).
\bibitem{2016_Kaw} K. Kawashima, T. Kinjo, T. Nishio, S. Ishida, H. Fujihisa, Y. Gotoh, K. Kihou, H. Eisaki, Y. Yoshida, and A. Iyo, Superconductivity in Fe-Based Compound EuAFe$_{4}$As$_{4}$ (A = Rb and Cs), J. Phys. Soc. Jpn. {\bf 85}, 064710 (2016).
\bibitem{2020_Ros} P. Rosa, Y. Xu, M. Rahn, J. Souza, S. Kushwaha, L. Veiga, A. Bombardi, S. Thomas, M. Janoschek, E. Bauer, M. Chan, Z. Wang, J. Thompson, N. Harrison, P. Pagliuso, A. Bernevig, and F. Ronning, Colossal magnetoresistance in a nonsymmorphic antiferromagnetic insulator, npj Quantum Mater. {\bf 5}, 52 (2020).
\bibitem{2023_May} A. H. Mayo, H. Takahashi, M. S. Bahramy, A. Nomoto, H. Sakai, and S. Ishiwata, Magnetic Generation and Switching of Topological Quantum Phases in a Trivial Semimetal ${\alpha}$-EuP$_{3}$, Phys. Rev. X {\bf 12}, 011033 (2023).
\bibitem{2009_Nan} S. Nandi, A. Kreyssig, Y. Lee, Y. Singh, J. W. Kim, D. C. Johnston, B. N. Harmon, and A. I. Goldman, Magnetic ordering in EuRh$_{2}$As$_{2}$ studied by x-ray resonant magnetic scattering, Phys. Rev. B {\bf 79}, 100407(R) (2009).\bibitem{2020_Tak} H. Takahashi, K. Aono, Y. Nambu, R. Kiyanagi, T. Nomoto, M. Sakano, K. Ishizaka, R. Arita, and S. Ishiwata, Competing spin modulations in the magnetically frustrated semimetal EuCuSb, Phys. Rev. B {\bf 102}, 174425 (2020).\bibitem{2022_Kur} T. Kurumaji, M. Gen, S. Kitou, H. Sagayama, A. Ikeda, and T. Arima, Anisotropic magnetotransport properties coupled with spiral spin modulation in a magnetic semimetal EuZnGe, Phys. Rev. Mater. {\bf 6}, 094410 (2022).
\bibitem{2023_Soh_2} J.-R. Soh, I. S.-Ram\'{i}rez, X. Yang, J. Sun, I. Zivkovic, J. A. R.-Velamaz\'{a}n, O. Fabelo, A. Stunault, A. Bombardi, C. Balz, M. D. Le, H. C. Walker, J. H. Dil, D. Prabhakaran, H. M. R\o nnow, F. de Juan, M. G. Vergniory, and A. T. Boothroyd, Weyl metallic state induced by helical magnetic order, npj Quantum Mater. {\bf 9}, 7 (2024).
\bibitem{2019_Kan} K. Kaneko, M. D. Frontzek, M. Matsuda, A. Nakao, K. Munakata, T. Ohhara, M. Kakihana, Y. Haga, M. Hedo, T. Nakama, and Y. \={O}nuki, Unique Helical Magnetic Order and Field-Induced Phase in Trillium Lattice Antiferromagnet EuPtSi, J. Phys. Soc. Jpn. {\bf 88}, 013702 (2019).\bibitem{2022_Tak} R. Takagi, N. Matsuyama, V. Ukleev, L. Yu, J. S. White, S. Francoual, J. R. L. Mardegan, S. Hayami, H. Saito, K. Kaneko, K. Ohishi, Y. \={O}nuki, T. Arima, Y. Tokura, T. Nakajima, and S. Seki, Nat. Commun. {\bf 13}, 1472 (2022).\bibitem{2023_Gen} M. Gen, R. Takagi, Y. Watanabe, S. Kitou, H. Sagayama, N. Matsuyama, Y. Kohama, A. Ikeda, Y. \={O}nuki, T. Kurumaji, T. Arima, and S. Seki, Rhombic skyrmion lattice coupled with orthorhombic structural distortion in EuAl$_{4}$, Phys. Rev. B {\bf 107}, L020410 (2023).
\bibitem{2023_Sin} D. Singh, Y. Fujishiro, S. Hayami, S. H. Moody, T. Nomoto, P. R. Baral, V. Ukleev, R. Cubitt, N.-J. Steinke, D. J. Gawryluk, E. Pomjakushina, Y. \={O}nuki, R. Arita, Y. Tokura, N. Kanazawa, and J. S. White, Transition between distinct hybrid skyrmion textures through their hexagonal-to-square crystal transformation in a polar magnet, Nat. Commun. {\bf 14}, 8050 (2023).
\bibitem{2023_Mat} T. Matsumura, K. Kurauchi, M. Tsukagoshi, N. Higa, H. Nakao, M. Kakihana, M. Hedo, T. Nakama, and Y. \={O}nuki, Helicity Unification by Triangular Skyrmion Lattice Formation in the Noncentrosymmetric Tetragonal Magnet EuNiGe$_{3}$, J. Phys. Soc. Jpn. {\bf 93}, 074705 (2024).
\bibitem{2006_Jia} J. Jiang and S. M. Kauzlarich, Colossal Magnetoresistance in a Rare Earth Zintl Compound with a New Structure Type: EuIn$_{2}$P$_{2}$, Chem. Mater. {\bf 18}, 435 (2006).
\bibitem{2019_Ma} J.-Z. Ma, S. M. Nie, C. J. Yi, J. Jandke, T. Shang, M. Y. Yao, M. Naamneh, L. Q. Yan, Y. Sun, A. Chikina, {\it et al}., Spin Fluctuation Induced Weyl Semimetal State in the Paramagnetic Phase of EuCd$_{2}$As$_{2}$, Sci. Adv. {\bf 5,} eaaw4718 (2019).
\bibitem{2020_Su} H. Su, B. Gong, W. Shi, H. Yang, H. Wang, W. Xia, Z. Yu. P.-J. Guo, J. Wang, L. Ding, L. Xu, X. Li, X. Wang, Z. Zou, N. Yu, Z. Zhu, Y. Chen, Z. Liu, K. Liu, G. Li, and Y. Guo, Magnetic exchange induced Weyl state in a semimetal EuCd$_{2}$Sb$_{2}$, APL Mater. {\bf 8}, 011109 (2020).
\bibitem{2021_Wan} Z.-C. Wang, J. D. Rogers, X. Yao, R. Nichols, K. Atay, B. Xu, J. Franklin, I. Sochnikov, P. J. Ryan, D. Haskel, and F. Taft, Colossal Magnetoresistance without Mixed Valence  in a Layered Phosphide Crystal, Adv. Mater. 33, 2005755 (2021).
\bibitem{2022_Bia} J. Blawat, M. Marshall, J. Singleton, E. Feng, H. Cao, W. Xie, and R. Jin, Unusual Electrical and Magnetic Properties in Layered EuZn$_{2}$As$_{2}$, Adv. Quantum Tech. {\bf 5}, 2200012 (2022).\bibitem{2023_Kre} S. Krebber, M. Kopp, C. Garg, K. Kummer, J. Sichelschmidt, S. Schulz, G. Poelchen, M. Mende, A. V. Virovets, K. Warawa, M. D. Thomson, A. V. Tarasov, D. Yu. Usachov, D. V. Vyalikh, H. G. Roskos, J. M\"{u}ller, C. Krellner, and K. Kliemt, Colossal magnetoresistance in EuZn$_{2}$P$_{2}$ and its electronic and magnetic structure, Phys. Rev. B {\bf 108}, 045116 (2023).\bibitem{2023_Cuo} G. Cuono, R. M. Sattigeri, C. Autieri, and T. Dietl, {\it Ab initio} overestimation of the topological region in Eu-based compounds, Phys. Rev. B {\bf 108}, 075150 (2023).
\bibitem{2009_Ess} A. M. Essin, J. E. Moore, and D. Vanderbilt, Magnetoelectric polarizability and axion electrodynamics in crystalline insulators, Phys. Rev. Lett. {\bf 102}, 146805 (2009).
\bibitem{2019_Xu} Y. Xu, Z. Song, Z. Wang, H. Weng, and X. Dai, Higher-Order Topology of the Axion Insulator EuIn$_{2}$As$_{2}$, Phys. Rev. Lett. {\bf 122}, 256402 (2019).\bibitem{2019_Tok} Y. Tokura, K. Yasuda, and A. Tsukazaki, Magnetic topological insulators, Nat. Rev. Phys. {\bf 1}, 126 (2019).
\bibitem{2022_Ber} B. A. Bernevig, C. Felser, and H. Beidenkopf, Progress and prospects in magnetic topological materials, Nature {\bf 603}, 41 (2022).\bibitem{2021_Fij} K. M. Fijalkowski, N. Liu, M. Hartl, M. Winnerlein, P. Mandal, A. Coschizza, A. Fothergill, S. Grauer, S. Schreyeck, K. Brunner, M. Greiter, R. Thomale, C. Gould, and L. W. Molenkamp, Any axion insulator must be a bulk three-dimensional topological insulator, Phys. Rev. B {\bf 103}, 235111 (2021). 
\bibitem{2011_Wan} X. Wan, A. M. Turner, A. Vishwanath, and S. Y. Savrasov, Topological semimetal and Fermi-arc surface states in the electronic structure of pyrochlore iridates, Phys. Rev. B {\bf 83}, 205101 (2011).\bibitem{2017_Mog} M. Mogi, M. Kawamura, R. Yoshimi, A. Tsukazaki, Y. Kozuka, N. Shirakawa, K. S. Takahashi, M. Kawasaki, and Y. Tokura, A magnetic heterostructure of topological insulators as a candidate for an axion insulator, Nat. Mater. {\bf 16}, 516 (2017).
\bibitem{2019_Zha} D. Zhang, M. Shi, T. Zhu, D. Xing, H. Zhang, and J. Wang, Topological Axion States in the Magnetic Insulator MnBi$_{2}$Te$_{4}$ with the Quantized Magnetoelectric Effect, Intrinsic magnetic topological insulators in van der Waals layered MnBi$_{2}$Te$_{4}$-family materials, Phys. Rev. Lett. {\bf 122}, 206401 (2019).
\bibitem{2019_Li} J. Li, Y. Li, S. Du, Z. Wang, B.-L. Gu, S.-C. Zhang, K. He, W. Duan, and Y. Xu, Sci. Adv, {\bf 5}, eaaw5685 (2019).
\bibitem{2020_Den} Y. Deng, Y. Yu, M. Z. Shi, Z. Guo, Z. Xu, J. Wang, X. H. Chen, Y. Zhang, Quantum anomalous Hall effect in intrinsic magnetic topological insulator MnBi$_{2}$Te$_{4}$, Science {\bf 367}, 895 (2020).
\bibitem{2008_Gof} A. M. Goforth, P. Klavins, J. C. Fettinger, and S. M. Kauzlarich, Magnetic Properties and Negative Colossal Magnetoresistance of the Rare Earth Zintl phase EuIn$_{2}$As$_{2}$, Inorg. Chem. {\bf 47}, 11048 (2008).
\bibitem{2020_Zha} Y. Zhang, K. Deng, X. Zhang, M. Wang, Y. Wang, C. Liu, J.-W. Mei, S. Kumar, E. F. Schwier, K. Shimada, C. Chen, and B. Shen, In-plane antiferromagnetic moments and magnetic polaron in the axion topological insulator candidate EuIn$_{2}$As$_{2}$, Phys. Rev. B {\bf 101}, 205126 (2020).
\bibitem{2020_Sat} T. Sato, Z. Wang, D. Takane, S. Souma, C. Cui, Y. Li, K. Nakayama, T. Kawakami, Y. Kubota, C. Cacho, T. K. Kim, A. Arab, V. N. Strocov, Y. Yao, and T. Takahashi, Signature of band inversion in the antiferromagnetic phase of axion insulator candidate EuIn$_{2}$As$_{2}$, Phys. Rev. Res. {\bf 2}, 033342 (2020).
\bibitem{2020_Reg} S. Regmi, M. M. Hosen, B. Ghosh, B. Singh, G. Dhakal, C. Sims, B. Wang, F. Kabir, K. Dimitri, Y. Liu, A. Agarwal, H. Lin, D. Kaczorowski, A. Bansil, and M. Neupane, Temperature-dependent electronic structure in a higher-order topological insulator candidate EuIn$_{2}$As$_{2}$, Phys. Rev. B {\bf 102}, 165153 (2020).
\bibitem{2020_Yu} F. H. Yu, H. M. Mu, W. Z. Zhuo, Z. Y. Wang, Z. F. Wang, J. J. Ying, and X. H. Chen, Elevating the magnetic exchange coupling in the compressed antiferromagnetic axion insulator candidate EuIn$_{2}$As$_{2}$, Phys. Rev. B {\bf 102}, 180404(R) (2020).
\bibitem{2022_Gon} M. Gong, D. Sar, J. Friedman, D. Kaczorowski, S. A. Razek, W.-C. Lee, and P. Aynajian, Surface state evolution induced by magnetic order in axion insulator candidate EuIn$_{2}$As$_{2}$, Phys. Rev. B {\bf 106}, 125156 (2022).
\bibitem{2022_Yan} J. Yan, Z. Z. Jiang, R. C. Xiao, W. J. Lu, W. H. Song, X. B. Zhu, X. Luo, Y. P. Sun, and M. Yamashita, Field-induced topological Hall effect in antiferromagnetic axion insulator candidate EuIn$_{2}$As$_{2}$, Phys. Rev. Res. {\bf 4}, 013163 (2022).
\bibitem{2023_Wu} Q. Wu, T. Hu, D. Wu, R. Li, L. Yue, S. Zhang, S. Xu, Q. Liu, T. Dong, and N. Wang, Spin dynamics in the axion insulator candidate EuIn$_{2}$As$_{2}$, Phys. Rev. B {\bf 107}, 174411 (2023).
\bibitem{2021_Rib} S. X. M. Riberolles, T. V. Trevisan, B. Kuthanazhi, T. W. Heitmann, F. Ye, D. C. Johnston, S. L. Bud'ko, D. H. Ryan, P. C. Canfield, A. Kreyssig, A. Vishwanath, R. J. McQueeney, L.-L. Wang, P. P. Orth, and B. G. Ueland, Magnetic crystalline-symmetry-protected axion electrodynamics and field-tunable unpinned Dirac cones in EuIn$_{2}$As$_{2}$, Nat. Commun. {\bf 12}, 999 (2021).
\bibitem{2023_Soh} J.-R. Soh, A. Bombardi, F. Mila, M. C. Rahn, D. Prabhakaran, S. Francoual, H. M.~R\o nnow, and A. T. Boothroyd, Understanding unconventional magnetic order in a candidate axion insulator by resonant elastic x-ray scattering, Nat. Commun. {\bf 14}, 3387 (2023).
\bibitem{2023_Don} E. Donoway, T. V. Trevisan, A. Liebman - Pel\'{a}ez, R. P. Day, K. Yamakawa, Y. Sun, J. R. Soh, D. Prabhakaran, A. Boothroyd, R. M. Fernandes, J. G. Analytis, J. E. Moore, J. Orenstein, and V. Sunko, Symmetry-breaking pathway towards the unpinned broken helix, arXiv:2310.16018.
\bibitem{2023_Cao} W. Cao, H. Yang, Y. Li, C. Pei, Q. Wang, Y. Zhao, C. Li, M. Zhang, S. Zhu, J. Wu, L. Zhang, Z. Wang, Y. Yao, Z. Liu, Y. Chen, and Y. Qi, Pressure-induced superconductivity in the Zintl topological insulator SrIn$_{2}$As$_{2}$, Phys. Rev. B {\bf 108}, 224510 (2023).
\bibitem{SM} See Supplemental Material at xxxx for details of the experiments and analyses, which includes Refs.~\cite{RIETAN, CrysAlis, Jana, 1948_Ise}.
\bibitem{RIETAN} F. Izumi and K. Momma, Three-Dimensional Visualization in Powder Diffraction, Solid State Phenom. {\bf 130}, 15 (2007).
\bibitem{CrysAlis} CrysAlisPro (Agilent Technologies Ltd, Yarnton, 2014).
\bibitem{Jana} V. Pet\v{r}\'{i}\v{c}ek, M. Du\v{s}ek, and L. Palatinus, Crystallographic computing system JANA2006: General features, Z. Kristallogr. Cryst. Mater. {\bf 229}, 345 (2014).
\bibitem{1948_Ise} I. Isenberg, B. R. Russell, and R. F. Greene, Improved Method for Measuring Hall Coefficients, Rev. Sci. Instrum. {\bf 19}, 685 (1948).
\bibitem{2003_Kos} T. Kosugia, S. Kawanob, N. Achiwac, A. Onoderad, Y. Nakaie, and N. Yamamoto, Direct evidence of helifan structures in holmium by single crystal neutron diffraction, Physica B {\bf 334}, 365 (2003).
\bibitem{2020_Ghi} N. J. Ghimire, R. L. Dally, L. Poudel, D. C. Jones, D. Michel, N. Thapa Magar, M. Bleuel, M. A. McGuire, J. S. Jiang, J. F. Mitchell, J. W. Lynn, I. I. Mazin, Competing magnetic phases and fluctuation-driven scalar spin chirality in the kagome metal YMn$_{6}$Sn$_{6}$, Sci. Adv. {\bf 6}, eabe2680 (2020).
\bibitem{comment} Note that the relation between $S({\mathbf q})/S({\mathbf Q}_{2})$ and $m_{{\mathbf Q}_{1}}/m_{{\mathbf Q}_{2}}$ does not depend on the $q_{1z}$ value.
\bibitem{2017_Hay} S. Hayami, R. Ozawa, and Y. Motome, Effective bilinear-biquadratic model for noncoplanar ordering in itinerant magnets, Phys. Rev. B {\bf 95}, 224424 (2017).
\end{thebibliography}
\end{document}


\title{Supplemental Material for\\``Incommensurate broken-helix induced by nonstoichiometry in the axion insulator candidate EuIn$_{2}$As$_{2}$''}

\author{Masaki~Gen}
\email{gen@issp.u-tokyo.ac.jp}
\affiliation{RIKEN Center for Emergent Matter Science (CEMS), Wako 351-0198, Japan}
\affiliation{Institute for Solid State Physics, University of Tokyo, Kashiwa, Chiba 277-8581, Japan}

\author{Yukako~Fujishiro}
\affiliation{RIKEN Center for Emergent Matter Science (CEMS), Wako 351-0198, Japan}

\author{Kazuki~Okigami}
\affiliation{Department of Applied Physics, University of Tokyo, Tokyo 113-8656, Japan}

\author{Satoru~Hayami}
\affiliation{Graduate School of Science, Hokkaido University, Sapporo 060-0810, Japan}

\author{Max T. Birch}
\affiliation{RIKEN Center for Emergent Matter Science (CEMS), Wako 351-0198, Japan}

\author{Kiyohiro~Adachi}
\affiliation{Materials Characterization Support Team, RIKEN Center for Emergent Matter Science (CEMS), Wako 351-0198, Japan}

\author{Daisuke~Hashizume}
\affiliation{Materials Characterization Support Team, RIKEN Center for Emergent Matter Science (CEMS), Wako 351-0198, Japan}

\author{Takashi~Kurumaji}
\affiliation{Department of Advanced Materials Science, University of Tokyo, Kashiwa 277-8561, Japan}
\affiliation{Division of Physics, Mathematics and Astronomy, California Institute of Technology, Pasadena, California 91125, USA}

\author{Hajime~Sagayama}
\affiliation{Institute of Materials Structure Science, High Energy Accelerator Research Organization, Tsukuba 305-0801, Japan}

\author{Hironori~Nakao}
\affiliation{Institute of Materials Structure Science, High Energy Accelerator Research Organization, Tsukuba 305-0801, Japan}

\author{Yoshinori~Tokura}
\affiliation{RIKEN Center for Emergent Matter Science (CEMS), Wako 351-0198, Japan}
\affiliation{Department of Applied Physics, University of Tokyo, Tokyo 113-8656, Japan}
\affiliation{Tokyo College, University of Tokyo, Tokyo 113-8656, Japan}

\author{Taka-hisa~Arima}
\affiliation{RIKEN Center for Emergent Matter Science (CEMS), Wako 351-0198, Japan}
\affiliation{Department of Advanced Materials Science, University of Tokyo, Kashiwa 277-8561, Japan}

\maketitle

\onecolumngrid

\vspace{+0.5cm}
This Supplemental Material includes contents as listed below:\\
\\
Note~1.~Single-crystal growth and characterization\\
Note~2.~Powder XRD experiment at low temperatures\\
Note~3.~Sample and strain dependence of RXS profiles at 5~K at zero field\\
Note~4.~Single-crystal XRD experiments and structural analyses\\
Note~5.~Hall resistivity measurements\\
Note~6.~Observation of higher-harmonic peaks\\
Note~7.~Polarization analysis for the RXS data at 5~K\\
Note~8.~Estimation of the amplitude of the ${\mathbf Q}_{1}$ and ${\mathbf Q}_{2}$ modulations\\
Note~9.~Theoretical calculations based on an effective spin Hamiltonian\\

\vspace{+1.5cm}
\subsection*{\label{Sec1} \large Note 1. Single-crystal growth and characterization}

Single crystals of EuIn$_{2}$As$_{2}$ were grown by an indium flux method.
The chunks of Eu, In, and As were weighed with a molar ratio of 3:36:9 (2.5--3.0~g total weight) and placed in alumina crucibles, which were then sealed in a fused silica tube.
The sample was heated up to 900~$^{\circ}$C over 1 day, followed by a 2 hours dwell time.
It was then cooled down to 770~$^{\circ}$C over 48 hours, at which point the excess indium flux was removed by a centrifuge.

The sample purity was checked by the powder x-ray diffraction (XRD) measurement on crushed single crystals at room temperature, using Rigaku SmartLab diffractometer at Materials Characterization Support Team, RIKEN CEMS.
The XRD patterns were recorded on a diffractometer with Bragg-Brentano geometry, and the incident x-ray beam was monochromatized by a Johansson-type monochromator with a Ge(111) crystal to select only Cu-$K\alpha_{1}$ radiation.
We confirm the hexagonal space group $P6_{3}/mmc$ (No.~194) for the main EuIn$_{2}$As$_{2}$ phase, in together with minor impurity phases of In [Space group $I4/mmm$ (No.~139)] and InAs [Space group $F{\overline 4}3m$ (No.~216)].
The Rietveld analysis was performed using the RIETAN software \cite{RIETAN}, yielding lattice parameters $a = b = 4.20521(4)~\AA$ and $c = 17.8915(2)~\AA$, in accord with Refs.~\cite{2008_Gof, 2020_Zha} (See Note~2 for details).
The results of the powder XRD at low temperatures are described in Note~2, and those of the single-crystal XRD are described in Note~4.
The magnetization was measured using a superconducting quantum interference device (MPMS, Quantum Design), confirming that the magnetization data are almost consistent with previous reports \cite{2020_Zha, 2020_Yu, 2023_Soh, 2021_Rib}, as shown in Figs.~2(d) and 3(a) in the main text.

\clearpage
\subsection*{\label{Sec2} \large Note 2. Powder XRD experiment at low temperatures}

\begin{figure}[h]
\centering
\renewcommand{\thefigure}{S\arabic{figure}}
\includegraphics[width=0.8\linewidth]{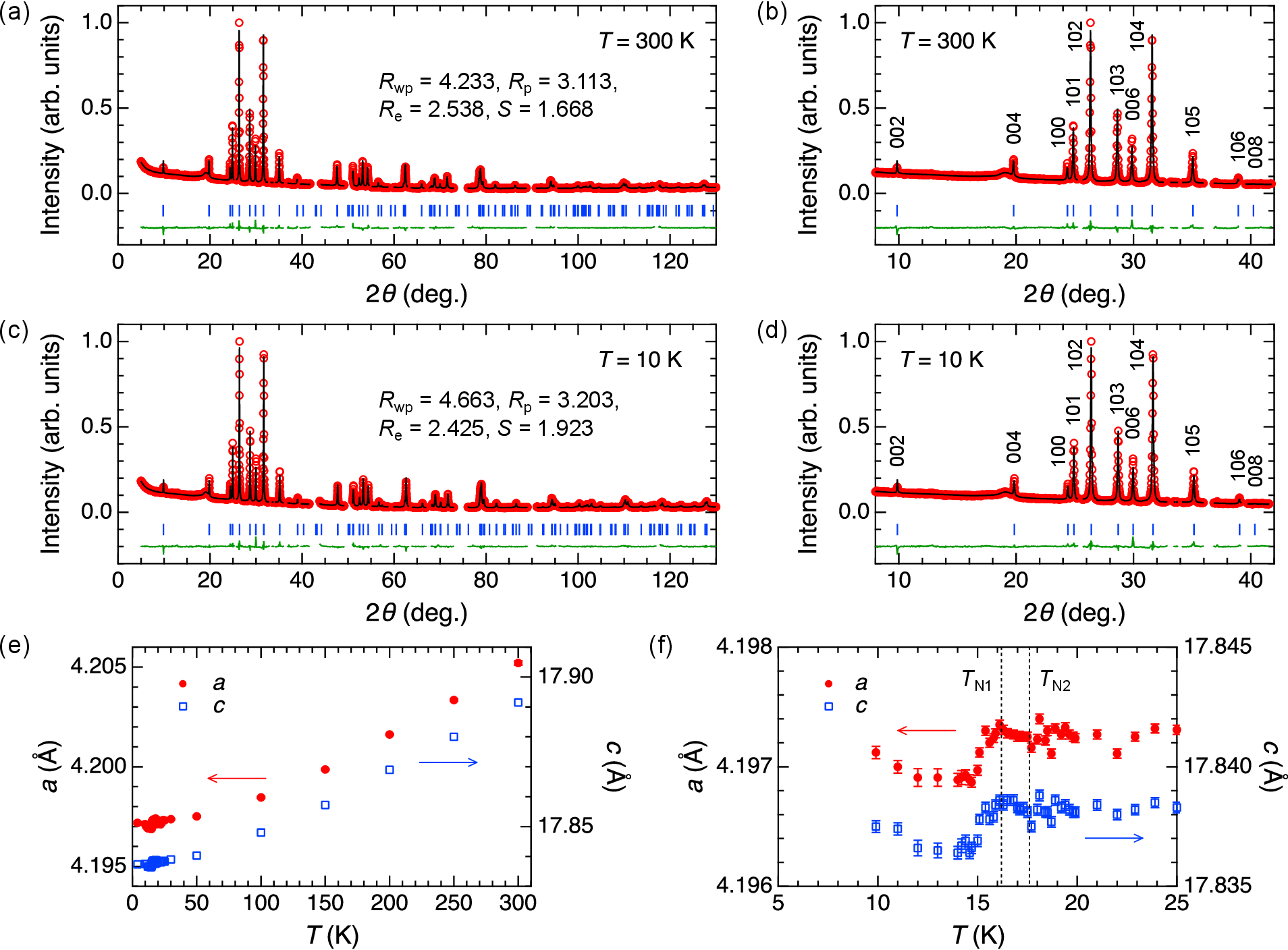}
\caption{[(a)--(d)] Powder XRD patterns and structural refinements of EuIn$_{2}$As$_{2}$ at (a) 300~K and (c) 10~K. Panels (b) and (d) display enlarged views of panels (a) and (b), respectively. The open red circles indicate the experimental data. The overplotted black curves indicate the calculated patterns with the hexagonal $P6_{3}/mmc$ space group as the main phase. Additional peaks originating from impurity phases and the background of a copper substrate were excluded ($2\theta \sim 42^{\circ}, 50^{\circ}, 55^{\circ}, 59^{\circ}, 63^{\circ}, 67^{\circ}, 75^{\circ}, 90^{\circ}, 95^{\circ}, 117^{\circ}$). The vertical bars indicate the positions of the Bragg reflections. The bottom curves show the difference between the experimental and calculated intensities. [(e)(f)] Temperature dependence of the lattice constants $a$ (left axis) and $c$ (right axis) obtained from the Rietveld analysis. Panel (f) is an enlarged view of panel (e) at low temperatures.}
\label{FigS1}
\end{figure}

Table~\ref{TabS1} and Fig.~\ref{FigS1} summarize the results of the powder XRD experiment performed between 10 and 300~K. 
Neither peak splitting nor the appearance of new peaks indicative of a structural phase transition was observed below the magnetic ordering temperature $T_{\rm N2} = 16.2$~K, as shown in Fig.~\ref{FigS1}(d).

\begin{table}[h]
\centering
\renewcommand{\arraystretch}{1.2}
\renewcommand{\thetable}{S\arabic{table}}
\caption{Structural parameters of EuIn$_{2}$As$_{2}$ at 300~K (top) and 10~K (bottom) obtained from the powder XRD measurements on crushed single crystals. The hexagonal $P6_{3}/mmc$ space group is assumed. Reliability factors are shown in the inset of Figs.~\ref{FigS1}(a) and \ref{FigS1}(c).}
\vspace{+0.1cm}
\begin{tabular}{ccccccc} \hline\hline
\multicolumn{7}{l}{~$T = 300$~K ($a = 4.20521(10)$~\AA, $c = 17.89150(32)$~\AA)}\\
~~Site~~ & ~Symmetry~ & ~Occupancy~ & $x$ & $y$ & $z$ & $U$ (\AA$^{2}$) \\ \hline
~~~Eu~~~ & ~~$2a$~~ & ~1~(fix)~ & ~~0~~ & ~~0~~ & ~~0~~ & ~~0.01543(55)~~~ \\
~~~In~~~ & ~~$4f$~~ & ~1~(fix)~ & ~~1/3~~ & ~~2/3~~ & ~~0.32668(7)~~ & ~~0.01763(40)~~~ \\
~~~As~~~ & ~~$4f$~~ & ~1~(fix)~ & ~~2/3~~ & ~~1/3~~ & ~~0.39276(11)~~ & ~~0.01577(53)~~~ \\ \hline\hline
\multicolumn{7}{l}{~$T = 10$~K ($a = 4.19712(5)$~\AA, $c = 17.83751(25)$~\AA)}\\
~~Site~~ & ~Symmetry~ & ~Occupancy~ & $x$ & $y$ & $z$ & $U$ (\AA$^{2}$) \\ \hline
~~~Eu~~~ & ~~$2a$~~ & ~1~(fix)~ & ~~0~~ & ~~0~~ & ~~0~~ & ~~0.00961(48)~~~ \\
~~~In~~~ & ~~$4f$~~ & ~1~(fix)~ & ~~1/3~~ & ~~2/3~~ & ~~0.32721(7)~~ & ~~0.00858(34)~~~ \\
~~~As~~~ & ~~$4f$~~ & ~1~(fix)~ & ~~2/3~~ & ~~1/3~~ & ~~0.39260(11)~~ & ~~0.00700(47)~~~ \\ \hline\hline
\end{tabular}
\label{TabS1}
\end{table}

\clearpage
\subsection*{\label{Sec3} \large Note 3. Sample and strain dependence of RXS profiles at 5~K at zero field}

\begin{figure}[h]
\centering
\renewcommand{\thefigure}{S\arabic{figure}}
\vspace{+0.5cm}
\includegraphics[width=0.9\linewidth]{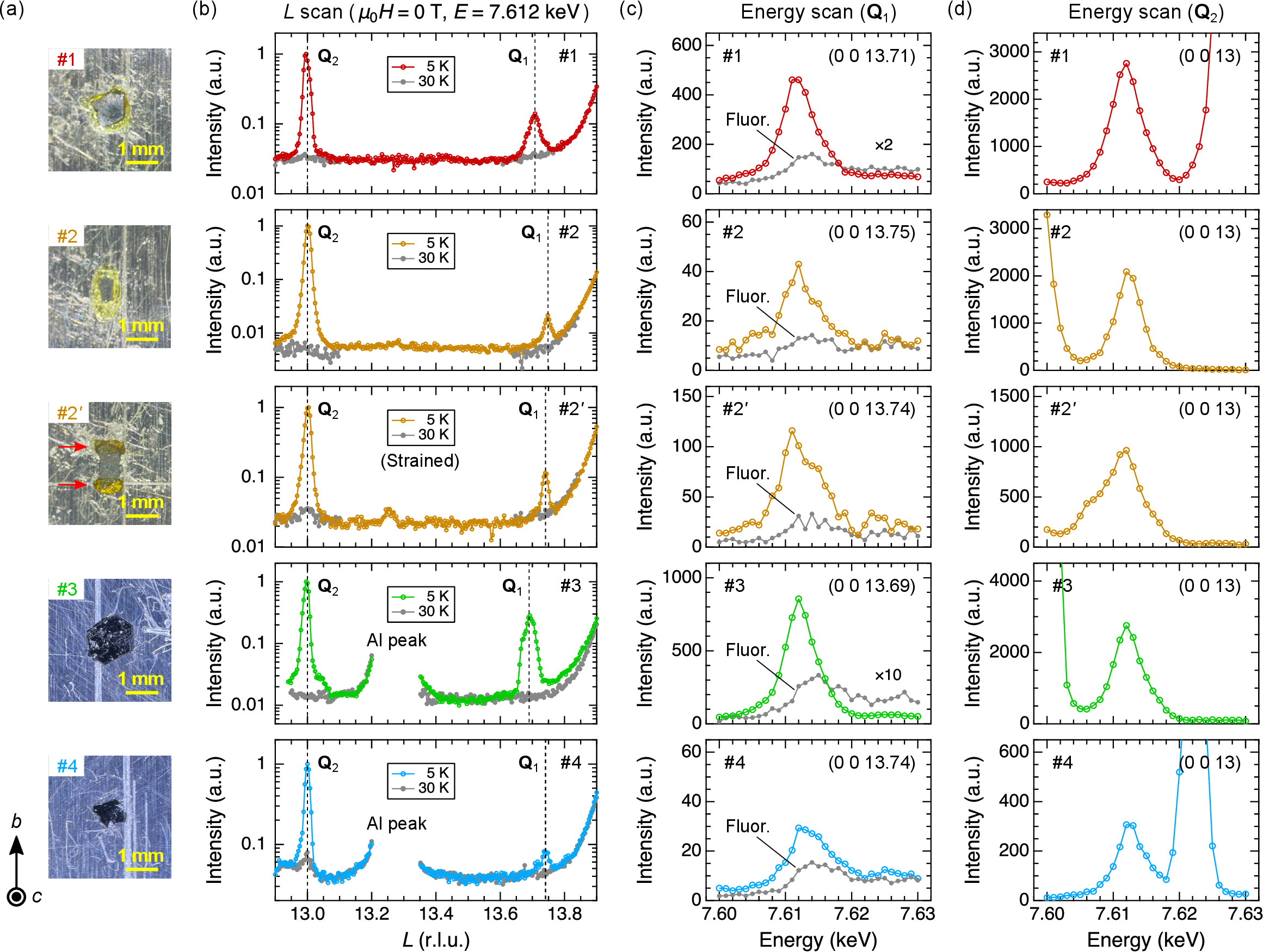}
\vspace{-0.1cm}
\caption{Results of RXS experiments for samples \#1, \#2, \#2' (strained), \#3, and \#4 from the top. (a) Photograph of EuIn$_{2}$As$_{2}$ single crystals, glued on an aluminum plate. (b) RXS profiles in the $(0, 0, L)$ scan at 5~K and 30~K at zero field. The left axis is plotted logarithmically. The data for sample \#1 is identical with that shown in Fig.~2(c) in the main text. [(c)(d)] Energy scan of magnetic Bragg spots at (c) $(0, 0, 14-q_{1z})$ and (d) (0, 0, 13). Energy scan of fluorescence is plotted in gray in panel (c). In panel (d), another intensity enhancement, besides the magnetic resonance at $E = 7.612$~keV, originates from multiple scattering.}
\label{FigS2}
\end{figure}

In order to investigate the sample dependence of the magnetic structure in EuIn$_{2}$As$_{2}$, we performed the resonant x-ray scattering (RXS) experiments on four distinct specimens (samples \#1 $\sim$ \#4). 
The four single crystals were picked up from the same batch, and the as-grown (001) surfaces were polished to remove the In flux.
In addition, we cut sample \#2 into two pieces (\#2 and \#2').
We glued a flat (001) plane on an aluminum plate homogeneously using GE vanish for samples \#1 $\sim$ \#4.
In order to verify the effect of extrinsic strain in our sample setting, we attached sample \#2' in another way, where only the two short sides facing each other along the $b$ axis were fixed to the Al plate using GE vanish, as indicated by red arrows in the middle panel of Fig.~\ref{FigS2}(a).
According to the powder XRD [Fig.~\ref{FigS1}(e)], the thermal shrinkage of the $ab$ plane of EuIn$_{2}$As$_{2}$ from room temperature to 10~K is approximately 0.2~\%, which is significantly small compared to that of Al ($\sim$0.7~\%).
Therefore, a compressive strain was expected to be applied along the $b$ axis for sample \#2'.

Figure~\ref{FigS2}(b) displays the RXS profiles obtained in the $(0, 0, L)$ scan at 5~K and 30~K at zero field for all the samples.
Here, the energy of the incident x-ray beam was $E=7.612$~keV, which was determined from the energy scan of the magnetic Bragg spots, as shown in Figs.~\ref{FigS2}(c) and \ref{FigS2}(d).
Two kinds of magnetic Bragg peaks, ${\mathbf Q}_{1} = (0, 0, q_{1z})$ and ${\mathbf Q}_{2} = (0, 0, 1)$, are commonly observed.
However, there are differences in the $q_{1z}$ value as well as the integrated-intensity ratio of the ${\mathbf Q}_{1}$ and ${\mathbf Q}_{2}$ peaks: $q_{1z} = 0.290$ and $I_{{\mathbf Q}_{1}}/I_{{\mathbf Q}_{2}} = 0.164(9)$ for sample \#1, $q_{1z} = 0.253$ and $I_{{\mathbf Q}_{1}}/I_{{\mathbf Q}_{2}} = 0.012(1)$ for sample \#2, $q_{1z} = 0.308$ and $I_{{\mathbf Q}_{1}}/I_{{\mathbf Q}_{2}} = 0.480(21)$ for sample \#3, and $q_{1z} = 0.260$ and $I_{{\mathbf Q}_{1}}/I_{{\mathbf Q}_{2}} = 0.039(6)$ for sample \#4.
Notably, the strained sample (\#2') exhibits $q_{1z} = 0.260$ and $I_{{\mathbf Q}_{1}}/I_{{\mathbf Q}_{2}} = 0.079(6)$, which are almost the same with those for sample \#2 with no strain.
From these results, we conclude that the observed variability in $q_{1z}$ and $I_{{\mathbf Q}_{1}}/I_{{\mathbf Q}_{2}}$ is primarily attributed to the sample dependence rather than extrinsic strain.
In Note~4, we show the evidence of a slight difference in the sample stoichiometry among the four samples.

\clearpage
\subsection*{\label{Sec4} \large Note 4. Single-crystal XRD experiments and structural analyses}

We performed single-crystal XRD measurements on four EuIn$_{2}$As$_{2}$ samples (\#1 $\sim$ \#4), which were obtained from the same batch and used for the RXS experiments.
The data were recorded on Rigaku MicroMax$^{\rm TM}$-007HF diffractometer equipped with a HyPix-6000 detector using Mo $K_{\alpha}$ radiation ($\lambda = 0.70926~\AA$) at room temperature.
The camera length was approximately 60~mm.
The intensities of Bragg reflections were collected by the CrysAlisPro software \cite{CrysAlis}.
Here, we employed absorption corrections based on the actual crystal shape for each sample in the finalization process.
The crystal structures were refined by the Jana2006 software \cite{Jana}.

All the structural parameters are summarized in Tables~\ref{TabS2}--\ref{TabS5}.
The crystallographic data are summarized in Table~\ref{TabS6}.
When optimizing the structural parameters, the occupancy of the Eu site was refined with fixing those of the In and As sites to 1.
As shown in Tables~\ref{TabS2}--\ref{TabS5}, the Eu deficiency is found to 1.64(46)\%, 1.10(39)\%, 1.42(36)\%, and 0.83(35)\% for samples \#1 $\sim$ \#4, respectively.
We confirm that there is no defect at In and As sites from the structure analysis with independently relaxing the occupancy of these sites.

Our structural analyses found differences in the lattice constants to be less than 0.1\% among the four samples.
Here, no clear correlation can be confirmed between the lattice constants and Eu deficiency, possibly due to the uncertainty in the actual camera length for each measurement (this uncertainty does not affect the analysis of Eu occupancy or the atomic positions).
We also note that the effect of stress would depend on the thermal expansion coefficient, not on the lattice constants themselves. 
Accordingly, the aforementioned minor differences in the lattice constants should have little effect on the strength of the uniaxial stress applied to the sample in our experimental setup.

\vspace{+0.5cm}
\begin{table}[H]
\renewcommand{\arraystretch}{1.2}
\renewcommand{\thetable}{S\arabic{table}}
\caption{Structural parameters of EuIn$_{2}$As$_{2}$ (sample \#1) at 300~K. The space group is $P6_{3}/mmc$ (No.~194), and the lattice parameters are $a = b = 4.210075(53)~\AA$, $c = 17.911096(254)~\AA$, $\alpha = \beta = 90^{\circ}$, $\gamma = 120^{\circ}$.}
\vspace{+0.1cm}
\hspace{+1.0cm}
\begin{tabular}{|c|c|c|c|c|c|c|c|c|} \hline
~~Site~~ & ~~Symmetry~~ & ~~Occupancy~~ & ~~$x$~~ & ~~$y$~~ & ~~$z$~~ & ~~$U_{11} (=U_{22})~(\AA^{2})$~~ & ~~$U_{33}~(\AA^{2})$~~ & ~~$U_{12}~(\AA^{2})$~~ \\ \hline
Eu & ~~$2a$~~ & ~~0.9836(46)~~ & ~~0~~ & ~~0~~ & ~~0~~ & ~~0.008504(138)~~ & ~~0.012646(213)~~ & ~~0.004252(69)~~ \\ \hline
In & ~~$4f$~~ & ~~1 (fix)~~ & ~~1/3~~ & ~~2/3~~ & ~~0.327193(26)~~ & ~~0.010401(139)~~ & ~~0.009377(174)~~ & ~~0.005200(70)~~ \\ \hline
As & ~~$4f$~~ & ~~1 (fix)~~ & ~~2/3~~ & ~~1/3~~ & ~~0.392478(39)~~ & ~~0.007285(172)~~ & ~~0.009782(256)~~ & ~~0.003643(86)~~ \\ \hline
\end{tabular}
\label{TabS2}
\end{table}

\begin{table}[H]
\renewcommand{\arraystretch}{1.2}
\renewcommand{\thetable}{S\arabic{table}}
\caption{Structural parameters of EuIn$_{2}$As$_{2}$ (sample \#2) at 300~K. The space group is $P6_{3}/mmc$ (No.~194), and the lattice parameters are $a = b = 4.210541(51)~\AA$, $c = 17.922931(227)~\AA$, $\alpha = \beta = 90^{\circ}$, $\gamma = 120^{\circ}$.}
\vspace{+0.1cm}
\hspace{+1.0cm}
\begin{tabular}{|c|c|c|c|c|c|c|c|c|} \hline
~~Site~~ & ~~Symmetry~~ & ~~Occupancy~~ & ~~$x$~~ & ~~$y$~~ & ~~$z$~~ & ~~$U_{11} (=U_{22})~(\AA^{2})$~~ & ~~$U_{33}~(\AA^{2})$~~ & ~~$U_{12}~(\AA^{2})$~~ \\ \hline
Eu & ~~$2a$~~ & ~~0.9890(39)~~ & ~~0~~ & ~~0~~ & ~~0~~ & ~~0.010279(98)~~ & ~~0.011567(130)~~ & ~~0.005140(49)~~ \\ \hline
In & ~~$4f$~~ & ~~1 (fix)~~ & ~~1/3~~ & ~~2/3~~ & ~~0.327150(19)~~ & ~~0.011983(99)~~ & ~~0.009045(109)~~ & ~~0.005991(49)~~ \\ \hline
As & ~~$4f$~~ & ~~1 (fix)~~ & ~~2/3~~ & ~~1/3~~ & ~~0.392463(29)~~ & ~~0.008805(121)~~ & ~~0.009645(165)~~ & ~~0.004402(61)~~ \\ \hline
\end{tabular}
\label{TabS3}
\end{table}

\begin{table}[H]
\renewcommand{\arraystretch}{1.2}
\renewcommand{\thetable}{S\arabic{table}}
\caption{Structural parameters of EuIn$_{2}$As$_{2}$ (sample \#3) at 300~K. The space group is $P6_{3}/mmc$ (No.~194), and the lattice parameters are $a = b = 4.211153(48)~\AA$, $c = 17.924691(160)~\AA$, $\alpha = \beta = 90^{\circ}$, $\gamma = 120^{\circ}$.}
\vspace{+0.1cm}
\hspace{+1.0cm}
\begin{tabular}{|c|c|c|c|c|c|c|c|c|} \hline
~~Site~~ & ~~Symmetry~~ & ~~Occupancy~~ & ~~$x$~~ & ~~$y$~~ & ~~$z$~~ & ~~$U_{11} (=U_{22})~(\AA^{2})$~~ & ~~$U_{33}~(\AA^{2})$~~ & ~~$U_{12}~(\AA^{2})$~~ \\ \hline
Eu & ~~$2a$~~ & ~~0.9858(36)~~ & ~~0~~ & ~~0~~ & ~~0~~ & ~~0.009926(99)~~ & ~~0.011354(118)~~ & ~~0.004963(50)~~ \\ \hline
In & ~~$4f$~~ & ~~1 (fix)~~ & ~~1/3~~ & ~~2/3~~ & ~~0.327139(17)~~ & ~~0.011793(100)~~ & ~~0.008763(100)~~ & ~~0.005896(50)~~ \\ \hline
As & ~~$4f$~~ & ~~1 (fix)~~ & ~~2/3~~ & ~~1/3~~ & ~~0.392460(27)~~ & ~~0.008538(123)~~ & ~~0.009275(150)~~ & ~~0.004269(61)~~ \\ \hline
\end{tabular}
\label{TabS4}
\end{table}

\begin{table}[H]
\renewcommand{\arraystretch}{1.2}
\renewcommand{\thetable}{S\arabic{table}}
\caption{Structural parameters of EuIn$_{2}$As$_{2}$ (sample \#4) at 300~K. The space group is $P6_{3}/mmc$ (No.~194), and the lattice parameters are $a = b = 4.208767(47)~\AA$, $c = 17.916377(265)~\AA$, $\alpha = \beta = 90^{\circ}$, $\gamma = 120^{\circ}$.}
\vspace{+0.1cm}
\hspace{+1.0cm}
\begin{tabular}{|c|c|c|c|c|c|c|c|c|} \hline
~~Site~~ & ~~Symmetry~~ & ~~Occupancy~~ & ~~$x$~~ & ~~$y$~~ & ~~$z$~~ & ~~$U_{11} (=U_{22})~(\AA^{2})$~~ & ~~$U_{33}~(\AA^{2})$~~ & ~~$U_{12}~(\AA^{2})$~~ \\ \hline
Eu & ~~$2a$~~ & ~~0.9917(35)~~ & ~~0~~ & ~~0~~ & ~~0~~ & ~~0.010228(96)~~ & ~~0.013535(153)~~ & ~~0.005114(48)~~ \\ \hline
In & ~~$4f$~~ & ~~1 (fix)~~ & ~~1/3~~ & ~~2/3~~ & ~~0.327123(20)~~ & ~~0.011956(97)~~ & ~~0.010925(125)~~ & ~~0.005978(48)~~ \\ \hline
As & ~~$4f$~~ & ~~1 (fix)~~ & ~~2/3~~ & ~~1/3~~ & ~~0.392449(31)~~ & ~~0.008794(118)~~ & ~~0.011102(186)~~ & ~~0.004397(59)~~ \\ \hline
\end{tabular}
\label{TabS5}
\end{table}

\begin{table}[t]
\renewcommand{\arraystretch}{1.2}
\renewcommand{\thetable}{S\arabic{table}}
\caption{Summary of crystallographic data of EuIn$_{2}$As$_{2}$ single crystals (samples \#1 $\sim$ \#4).}
\vspace{+0.1cm}
\begin{tabular}{|l|l|l|l|l|} \hline
~~~~ & ~~Sample \#1~~ & ~~Sample \#2~~ & ~~Sample \#3~~ & ~~Sample \#4~~ \\ \hline
~~Temperature~(K)~~ & ~~300~~ & ~~300~~ & ~~300~~ & ~~300~~ \\ \hline
~~Wavelength~($\AA$)~~ & ~~0.70926~~ & ~~0.70926~~ & ~~0.70926~~ & ~~0.70926~~ \\ \hline
~~Crystal dimension~($\mu$m$^{3}$)~~ & ~~$190 \times 140 \times 60$~~ & ~~$250 \times 150 \times 50$~~ & ~~$180 \times 140 \times 80$~~ & ~~$120 \times 70 \times 30$~~ \\ \hline
~~Space group~~ & ~~$P6_{3}/mmc$~~ & ~~$P6_{3}/mmc$~~ & ~~$P6_{3}/mmc$~~ & ~~$P6_{3}/mmc$~~ \\ \hline
~~$a~(\AA)$~~ & ~~4.210075(53)~~ & ~~4.210541(51)~~ & ~~4.211153(48)~~ & ~~4.208767(47)~~ \\ \hline
~~$c~(\AA)$~~ & ~~17.911096(254)~~~~~~~ & ~~17.922931(227)~~~~~~~ & ~~17.924691(160)~~~~~~~ & ~~17.916377(265)~~~~~~~ \\ \hline
~~$Z$~~ & ~~2~~ & ~~2~~ & ~~2~~ & ~~2~~ \\ \hline
~~$F(000)$~~ & ~~451.93~~ & ~~452.61~~ & ~~452.21~~ & ~~452.94~~ \\ \hline
~~$(\sin \theta/\lambda)_{\rm max}~(\AA^{-1})$~~ & ~~1.28~~ & ~~1.28~~ & ~~1.28~~ & ~~1.28~~ \\ \hline
~~$N_{\rm Total}$~~ & ~~26364~~ & ~~25288~~ & ~~27309~~ & ~~29166~~ \\ \hline
~~$N_{\rm Unique}$~~ & ~~970~~ & ~~1003~~ & ~~1000~~ & ~~1001~~ \\ \hline
~~Average redundancy~~ & ~~27.179~~ & ~~25.212~~ & ~~27.309~~ & ~~29.137~~ \\ \hline
~~Completeness~(\%)~~ & ~~96.9~~ & ~~100~~ & ~~99.9~~ & ~~100~~ \\ \hline
~~$N_{\rm parameter}$~~ & ~~11~~ & ~~11~~ & ~~11~~ & ~~11~~ \\ \hline
~~$R_{1}$~($I > 3\sigma$)~[number of reflections]~~~~~~~ & ~~3.19\%~[806]~~ & ~~2.61\%~[868]~~ & ~~2.80\%~[866]~~ & ~~2.71\%~[832]~~ \\ \hline
~~$R_{1}$~(all)~[number of reflections]~~ & ~~4.27\%~[970]~~ & ~~3.52\%~[1003]~~ & ~~3.70\%~[1000]~~ & ~~3.66\%~[1001]~~ \\ \hline
~~$wR_{2}$~(all)~[number of reflections]~~ & ~~8.68\%~[970]~~ & ~~6.63\%~[1003]~~ & ~~6.37\%~[1000]~~ & ~~6.36\%~[1001]~~ \\ \hline
~~GOF~(all)~[number of reflections]~~ & ~~4.23\%~[970]~~ & ~~3.80\%~[1003]~~ & ~~3.93\%~[1000]~~ & ~~3.58\%~[1001]~~ \\ \hline
\end{tabular}
\label{TabS6}
\end{table}

\clearpage
\subsection*{\label{Sec5} \large Note 5. Hall resistivity measurements}

As discussed in Notes.~3 and 4, we reveal that the magnetic structure of the EuIn$_{2}$As$_{2}$ single crystal exhibits variability in the ${\mathbf Q}_{1}$ modulation across samples, which is most likely caused by the difference in the amount of Eu deficiency.
Therefore, the carrier number as well as the Fermi energy are also expected to have sample dependence.
To verify this, we performed the Hall resistivity measurements for samples \#3 ($q_{1z} = 0.308$) and \#4 ($q_{1z} = 0.260$).

The transverse resistivity was measured at a frequency of 511~Hz using a lock-in amplifier (SR860, Stanford Research Systems).
The measurements were performed in a temperature range from 300~K to 2~K and in a magnetic field range from $-9$~T to 9~T using a commercial cryostat equipped with a superconducting magnet (physical property measurement system, Quantum Design).
The obtained transverse resistivities were field-antisymmetrized to correct contact misalignment.
For evaluating the Hall voltage, we corrected the measured voltage based on the crystal shape, according to Fig.~2 in Ref.~\cite{1948_Ise}. 

\begin{figure}[h]
\centering
\renewcommand{\thefigure}{S\arabic{figure}}
\vspace{+0.5cm}
\includegraphics[width=0.6\linewidth]{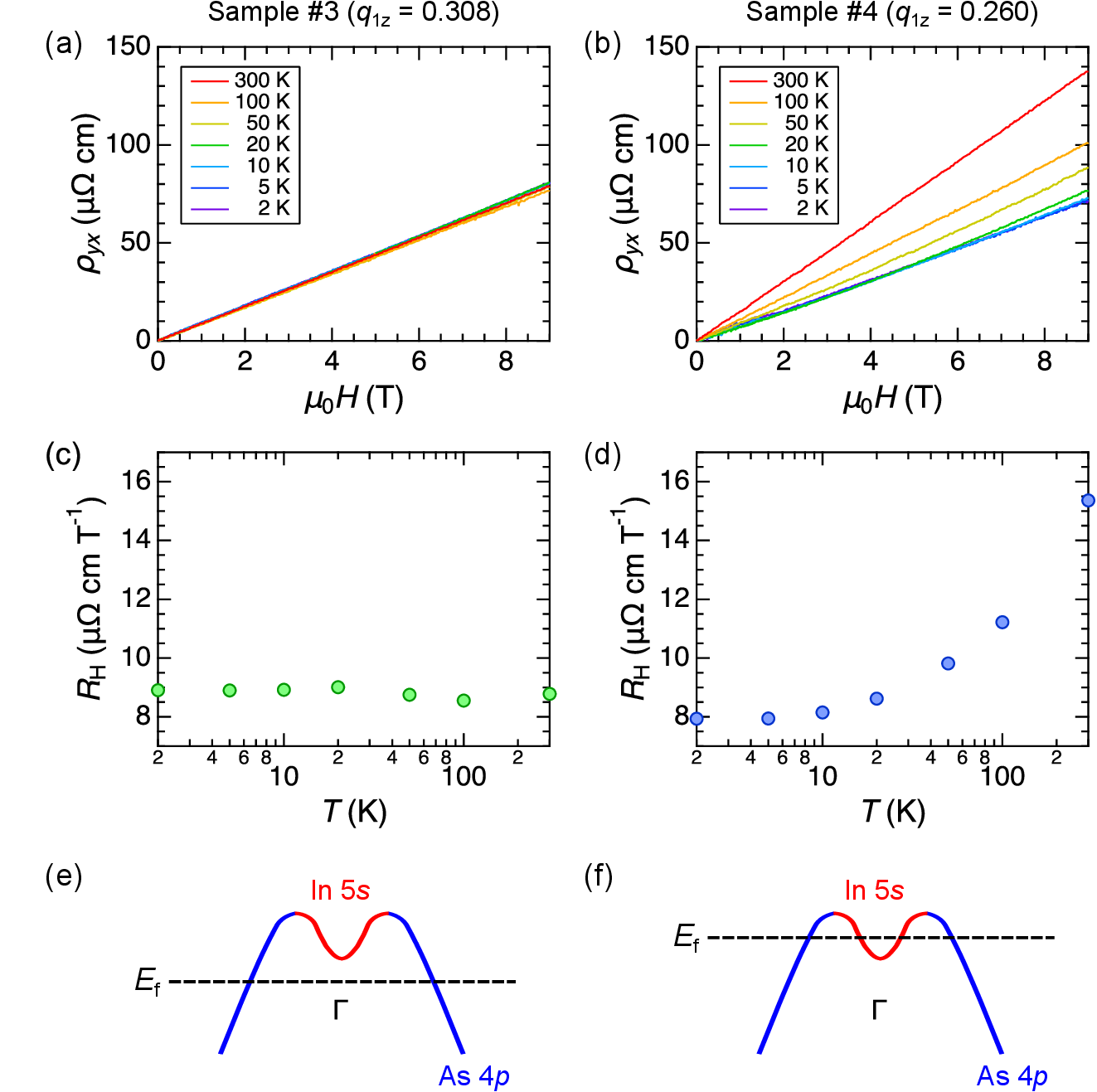}
\caption{[(a)(b)] Magnetic-field dependence of the Hall resistivity $\rho_{yx}$ at various temperatures for samples \#3 (a) and \#4 (b). [(c)(d)] Temperature dependence of the Hall coefficient $R_{\rm H}$ estimated assuming the single-band model for samples \#3 (c) and \#4 (d). [(e)(f)] Schematic of the electronic band structure for samples \#3 (e) and \#4 (f).}
\label{FigS3}
\end{figure}

Figures~\ref{FigS3}(a) and \ref{FigS3}(b) show the field dependence of the Hall resistivity $\rho_{yx}$ measured at various temperatures for samples \#3 and \#4, respectively.
For both samples, the $\rho_{yx}$--$H$ curves exhibit an overall positive slope, indicating dominant hole-carrier contributions at all measured temperatures.
By performing a linear fit to the $\rho_{yx}$--$H$ curve between 0 and 9~T for simplicity, we estimate the Hall coefficient $R_{\rm H}$ at each temperature, as shown in Figs.~\ref{FigS3}(c) and \ref{FigS3}(d).
Remarkably, $R_{\rm H}$ for sample \#4 dramatically increases with rising temperature.
This behavior cannot be explained by the temperature dependence of carrier density expected in ordinary semiconductors and strongly suggests a multi-carrier nature.
We also note that the $\rho_{yx}$--$H$ curves for sample \#4 exhibits a slight concave shape, indicative of the multi-carrier nature.
The Fermi level is expected to cross the In $5s$ band, as schematically illustrated in Fig.~\ref{FigS3}(f).
In contrast, $R_{\rm H}$ for sample \#3 shows almost no temperature dependence, suggesting little contribution of electron carriers to the electric transport.
The Fermi level would be located deeper in the As $4p$ band, as shown Fig.~\ref{FigS3}(e).
The above electronic band picture is consistent with a larger Eu deficiency for sample \#3 (1.42(36)\%) than for sample \#4 (0.83(35)\%).

Assuming a single-band model, the carrier number for sample \#3 at 2~K can be roughly estimated to $n \approx 7.2 \times 10^{-19}$~cm$^{-3}$, which is consistent with the previous reports \cite{2022_Yan}.
On the other hand, due to the multi-carrier nature, the carrier number for sample \#4 cannot be directly estimated solely from the Hall resistivity data.

\clearpage
\subsection*{\label{Sec6} \large Note 6. Observation of higher-harmonic peaks}

\begin{figure}[h]
\centering
\renewcommand{\thefigure}{S\arabic{figure}}
\vspace{+1.0cm}
\includegraphics[width=0.8\linewidth]{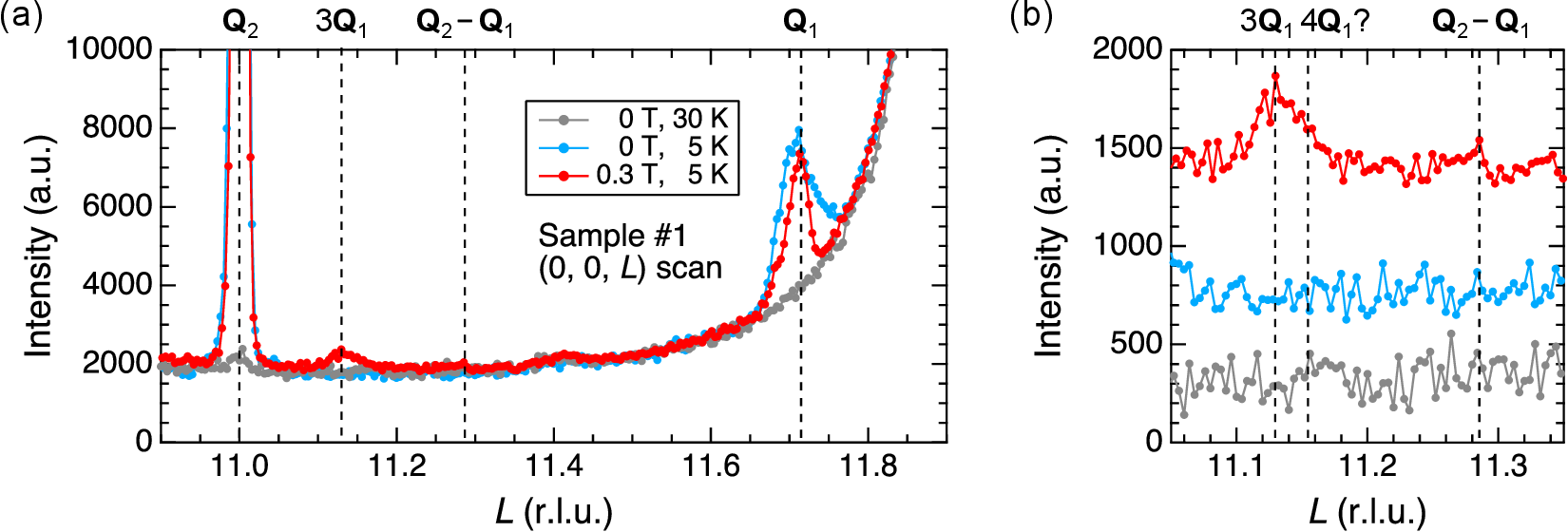}
\caption{(a) RXS profiles observed in the $(0, 0, L)$ scan with $L = 10.9$--11.9 for sample \#1 at 0~T and 30~K (gray), 0~T and 5~K (cyan), and 0.3~T and 5~K (red). The vertical dashed lines represent the positions of observed magnetic Bragg peaks. (b) Enlarged view of panel (a). The data are offset for clarity.}
\label{FigS4}
\end{figure}

Figure~\ref{FigS4}(a) shows the RXS profiles observed in the $(0, 0, L)$ scan at several temperature and magnetic-field conditions.
These data sets were obtained from scans different from those shown in Fig.~3 in the main text.
At 0~T and 5~K, the ${\mathbf Q}_{1}$ peak appears at $L \approx 11.71$, corresponding to $q_{1z} \approx 0.29$.
Furthermore, at 0.3~T, higher-harmonic peaks appear at $L \approx 11.13$ and 11.29, which likely correspond to $3{\mathbf Q}_{1}$ and ${\mathbf Q}_{2} - {\mathbf Q}_{1}$, respectively.
These high-harmonic peaks are absent in the paramagnetic phase at 0~T and 30~K, indicating that they are of magnetic origin.

As shown in the enlarged view in Fig.~\ref{FigS4}(b), the observed high-harmonic peak at $L \approx 11.13$ at 0.3~T and 5~K is broadened toward larger $L$.
The shoulder around $L \approx 11.16$ may be attributed to the $4{\mathbf Q}_{1}$ peak, which can be considered as a satellite peak of the fundamental Bragg peak at $L = 10$ (note that $10 + 0.29 \times 4 = 11.16$).
The superposition of ${\mathbf Q}_{1}$ and $3{\mathbf Q}_{1}$ may be the origin for the appearance of the $4{\mathbf Q}_{1}$ modulation.
Since it is difficult to resolve the $3{\mathbf Q}_{1}$ and the possible $4{\mathbf Q}_{1}$ peaks in our experimental data, we would like to simply refers to the observed higher-harmonic peak as originating from the $3{\mathbf Q}_{1}$ component in the main text.

\clearpage
\subsection*{\label{Sec7} \large Note 7. Polarization analysis for the RXS data at 5~K}

In order to reveal the orientation of the ${\mathbf Q}_{1}$ and ${\mathbf Q}_{2}$ modulations in the zero-field low-temperature (LT) phase below $T_{\rm N1} = 16.2$~K and the high-field (HF) phase above $\mu_{0}H_{\rm c1} = 0.2$~T with $H \parallel b$, we performed polarization analysis at 5~K for sample \#1 using incident x-rays in resonance with the Eu $L_{2}$ absorption edge ($E = 7.612$~keV).
As described in the main text, the magnetic scattering intensity $I$ is given by $I \propto ({\mathbf e}_{\rm i} \times {\mathbf e}_{\rm f}) \cdot {\mathbf m}_{\mathbf Q}$, where ${\mathbf e}_{\rm i}$ (${\mathbf e}_{\rm f}$) is the polarization vector of the incident (scattered) beam, and ${\mathbf m}_{\mathbf Q}$ is a spin moment for the ${\mathbf Q}$  modulation.
The $\pi$--$\pi'$ channel ($I_{\pi-\pi'}$) always detects the modulated component along $b$ ($m_{b}$), and the $\pi$--$\sigma'$ channel ($I_{\pi-\sigma'}$) detects those along $a^{*}$ ($m_{a^{*}}$) and $c$ ($m_{c}$).
For the latter, the intensity ratio of $m_{a^{*}}$ and $m_{c}$ is given by $\cos^{2}\omega : \sin^{2}\omega$, where $\omega$ corresponds to the angle between the propagation vector of the incident beam (${\mathbf k}_{\rm i}$) and the $a^{*}$ axis.
We focused on the magnetic Bragg spots at $(-2, 0, 11.71)$ and $(2, 0, 11.71)$ for analyzing the ${\mathbf Q}_{1}$ modulation and those at $(0, 0, 9)$ and $(0, 0, 17)$ for the ${\mathbf Q}_{2}$ modulation.

\begin{figure}[h]
\centering
\renewcommand{\thefigure}{S\arabic{figure}}
\vspace{+0.5cm}
\includegraphics[width=0.95\linewidth]{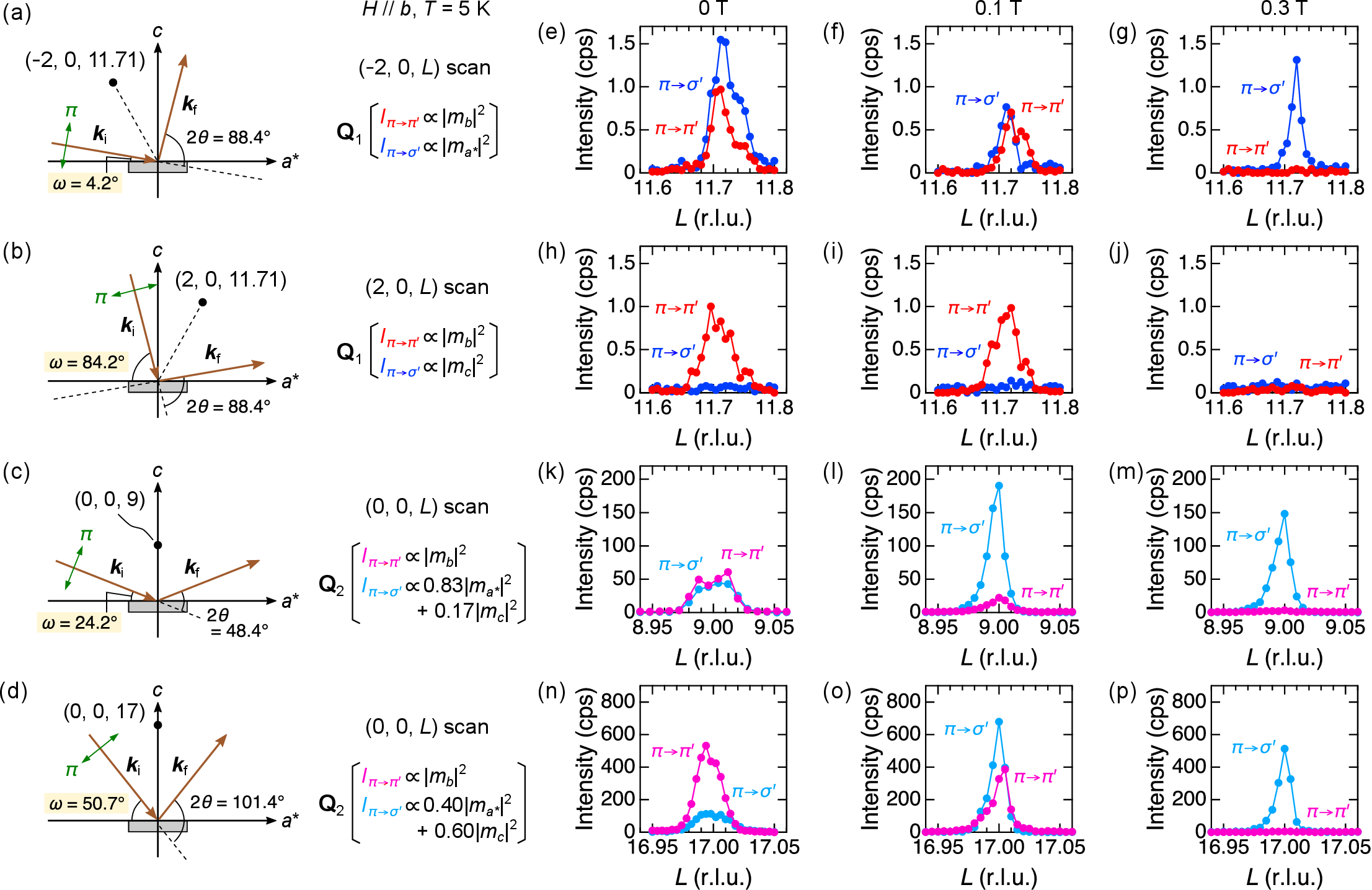}
\caption{[(a)--(d)] Experimental configuration of the RXS focusing on the magnetic Bragg spots at (a) $(-2, 0, 11.71)$, (b) $(2, 0, 11.71)$, (c) $(0, 0, 9)$, and (d) $(0, 0, 17)$. The former (latter) two correspond to the ${\mathbf Q}_{1}$ (${\mathbf Q}_{2}$) peak. [(e)--(p)] RXS profiles of the polarization analysis at 5~K at 0~T [(e)(h)(k)(n)], 0.1~T [(f)(i)(l)(o)], and 0.3~T [(g)(j)(m)(p)] with $H \parallel b$. Panels (e)--(g), (h)--(j), (k)--(m), and (n)--(p) show the data of the $(-2, 0, L)$ scan around $L = 11.71$, $(2, 0, L)$ scan around $L = 11.71$, $(0, 0, L)$ scan around $L = 9$, and $(0, 0, L)$ scan around $L = 17$, respectively.}
\label{FigS5}
\end{figure}

Notably, $\omega$ is close to $0^{\circ}$ and $90^{\circ}$ for the $(-2, 0, 11.71)$ and $(2, 0, 11.71)$ positions, respectively.
We can hence independently extract information on $m_{a^{*}}$ and $m_{c}$ of the ${\mathbf Q}_{1}$ modulation by measuring $I_{\pi-\sigma'}$ at $(-2, 0, 11.71)$ and $(2, 0, 11.71)$, respectively [Figs.~\ref{FigS5}(a) and \ref{FigS5}(b)].
As shown in Figs.~\ref{FigS5}(h)--\ref{FigS5}(j), $I_{\pi-\sigma'}$ exhibits no intensity in all the RXS profiles, indicating the absence of $m_{c}$ of the ${\mathbf Q}_{1}$ modulation in both the LT and HF phases.
As for the ${\mathbf Q}_{2}$ peaks, $\omega$ is $24.2^{\circ}$ for $(0, 0, 9)$ and $50.7^{\circ}$ for $(0, 0, 17)$.
Accordingly, 83~\% of $m_{a^{*}}$ and 17~\% of $m_{c}$ components contribute to $I_{\pi-\sigma'}$ at $(0, 0, 9)$, and 40~\% of $m_{a^{*}}$ and 60~\% of $m_{c}$ components contribute to $I_{\pi-\sigma'}$ at $(0, 0, 17)$.
At 0~T, the observed intensity ratio between the $\pi$--$\pi'$ and $\pi$--$\sigma'$ channels is $I_{\pi-\sigma'}/I_{\pi-\pi'} \approx 0.8$ at $(0, 0, 9)$ [Fig.~\ref{FigS5}(k)] and $I_{\pi-\sigma'}/I_{\pi-\pi'} \approx 0.3$ at $(0, 0, 17)$ [Fig.~\ref{FigS5}(n)].
The difference in $I_{\pi-\sigma'}/I_{\pi-\pi'}$ should be attributed to the difference in the contribution from $m_{a^{*}}$ of the ${\mathbf Q}_{2}$ modulation.
In other words, $m_{c}$ of the ${\mathbf Q}_{2}$ modulation should be absent in the LT phase.
This tendency also holds at 0.1~T: $I_{\pi-\sigma'}/I_{\pi-\pi'} \approx 7$ at $(0, 0, 9)$ [Fig.~\ref{FigS5}(l)] and $I_{\pi-\sigma'}/I_{\pi-\pi'} \approx 1.5$ at $(0, 0, 17)$ [Fig.~\ref{FigS5}(o)].
After the metamagnetic transition, $I_{\pi-\sigma'}$ have no intensity at both $(0, 0, 9)$ and $(0, 0, 17)$, indicating the absence of $m_{c}$ as well as $m_{a^{*}}$ of the ${\mathbf Q}_{2}$ modulation in the HF phase.
From the above considerations, we can conclude that all the spins lie within the $ab$ plane in the magnetic ordering phases in EuIn$_{2}$As$_{2}$.

\clearpage
\subsection*{\label{Sec8} \large Note 8. Estimation of the amplitude of the ${\mathbf Q}_{1}$ and ${\mathbf Q}_{2}$ modulations}

In Fig.~1(c) of the main text, we compare the integrated-intensity ratio of the ${\mathbf Q}_{1}$ and ${\mathbf Q}_{2}$ peaks, $I_{{\mathbf Q}_{1}}/I_{{\mathbf Q}_{2}}$, among six samples; four (samples \#1 $\sim$ \#4) were measured in our RXS experiments, and the other two were measured in the previous RXS experiment by Soh {\it et al.} \cite{2023_Soh} and the neutron diffraction (ND) experiment by Riberolles {\it et al.} \cite{2021_Rib}.
Also, we show that the broken-helix state can be approximately described by ${\mathbf S}_{i}$ = ${\mathbf m}({\mathbf r}_{i})/|{\mathbf m}({\mathbf r}_{i})|$, where ${\mathbf m}({\mathbf r}) \propto (1, i, 0) \sum_{\eta = 1, 2}{m_{{\mathbf Q}_{\eta}}}\exp(i{\mathbf Q}_{\eta}\cdot{\mathbf r}) + {\rm c.c.}$ at zero field.
In this section, we present the analysis method for the correction of $I_{{\mathbf Q}_{1}}/I_{{\mathbf Q}_{2}}$ and the estimation of $m_{{\mathbf Q}_{1}}/m_{{\mathbf Q}_{2}}$  for each sample.\\

\hspace{-0.35 cm}\underline{Our RXS data}\\

Here, we describe themethod for estimating $m_{{\mathbf Q}_{1}}/m_{{\mathbf Q}_{2}}$ in sample~\#1 ($q_{1z} = 0.290$) as an example.
First, we estimate the intensity ratio of the $00L$ reflections with $L = 13.71$ and $L = 13$ in the RXS profile at 5~K at zero field [see the top panel in Fig.~S2(b)].
From the Gaussian fit on each peak, we obtain $I_{{\mathbf Q}_{1}}/I_{{\mathbf Q}_{2}} = 0.164(9)$.
Since the magnetic scattering intensity is given by $I \propto |({\mathbf e}_{\rm i} \times {\mathbf e}_{\rm f}) \cdot {\mathbf m}_{\mathbf Q}|^{2}$, a ${\mathbf Q}$-dependent correction factor needs to be applied to the observed peak intensity.
As mentioned in Note~6, the $\pi-\pi'$ channel detects the $m_{b}$ component, while the $\pi-\sigma'$ channel detects the $m_{a^{*}}$ component (the absence of $m_{c}$ is confirmed).
Assuming that magnetic domains of the broken helix are randomly oriented within the $ab$ plane, the correction factor can be approximated to $(\sin^{2}2\theta +\cos^{2}\theta)/2$.
In our RXS experiments, the energy of the incident x-ray was $E = 7.612$~keV, in resonance with the Eu $L_{2}$ edge, so that $2\theta = 77.49^{\circ}$ for $L = 13.71$ and $2\theta = 72.81^{\circ}$ for $L = 13$.
Then, $(\sin^{2}2\theta +\cos^{2}\theta)/2$ is 0.781 for $L = 13.71$ and 0.780 for $L = 13$, i.e., the polarization effects on estimating $I_{{\mathbf Q}_{1}}/I_{{\mathbf Q}_{2}}$ are negligibly small.
Using these factors, the corrected intensity ratio is $I_{{\mathbf Q}_{1}}/I_{{\mathbf Q}_{2}} = 0.164(9)$.
Finally, we can estimate $m_{{\mathbf Q}_{1}}/m_{{\mathbf Q}_{2}}$ using the relation shown in Fig.~5(b) in the main text, yielding $m_{{\mathbf Q}_{1}}/m_{{\mathbf Q}_{2}} \approx 0.703(3)$.

$m_{{\mathbf Q}_{1}}/m_{{\mathbf Q}_{2}}$ for samples \#2 $\sim$ \#4 can be estimated by the same procedure as mentioned above, yielding $m_{{\mathbf Q}_{1}}/m_{{\mathbf Q}_{2}} \approx 0.220(1)$, 0.966(6), and 0.378(25) for samples \#2 ($q_{1z} = 0.235$), \#3 ($q_{1z} = 0.308$), and \#4 ($q_{1z} = 0.260$), respectively.\\

\hspace{-0.35 cm}\underline{RXS data by Soh {\it et al.}}\\

From the RXS profile at 6~K at zero field shown in Fig.~1(a) of Ref.~\cite{2023_Soh}, we estimate the intensity ratio of the $00L$ reflections with $L = 13\frac{2}{3}$ and $L = 13$ to be 1:1.24. 
Importantly, as the observed $q_{1z}$ value is 0.333 (close to 1/3), the higher-harmonic ${{\mathbf Q}_{2}}-2{{\mathbf Q}_{1}}$ and $3{{\mathbf Q}_{1}}$ peaks are expected to overlap with the ${{\mathbf Q}_{1}}$ and ${{\mathbf Q}_{2}}$ peaks, respectively.
Therefore, one needs to take into account the higher-harmonic contributions in estimating $m_{{\mathbf Q}_{1}}/m_{{\mathbf Q}_{2}}$.
For evaluating the correction factors, we here consider only the ${{\mathbf Q}_{1}}$ and ${{\mathbf Q}_{2}}$ peaks for simplicity.
Since the energy of the incident x-ray was $E = 6.975$~keV, in resonance with the Eu $L_{3}$ edge, $2\theta = 85.83^{\circ}$ for $L = 13\frac{2}{3}$ and $2\theta = 80.74^{\circ}$ for $L = 13$ (We here use the lattice constant $c = 17.83751(25)~\AA$ at 10~K obtained in our powder XRD measurement, as shown in Table~\ref{TabS1}.
Therefore, $(\sin^{2}2\theta +\cos^{2}\theta)/2$ is 0.766 for $L = 13.71$ and 0.778 for $L = 13$.
Using these correction factors and the relation shown in Figs.~5(b) in the main text, we finally obtain $m_{{\mathbf Q}_{1}}/m_{{\mathbf Q}_{2}} \approx 1.14$.\\

\hspace{-0.35 cm}\underline{ND data by Riberolles {\it et al.}}\\

In Supplementary Table III of Ref.~\cite{2021_Rib}, the authors provide intensity data for several magnetic peaks after absorption corrections.
The ND experiment detects magnetic moments perpendicular to the magnetic propagation vectors.
Considering that there is no magnetic moment along the $c$ axis in EuIn$_{2}$As$_{2}$, it is straightforward to compare the $00L$ magnetic reflections with each other because only the magnetic form factor $f({\mathbf Q}_{\eta})$ affects the magnetic scattering intensity.
Therefore, we choose six magnetic peaks for estimating $m_{{\mathbf Q}_{1}}/m_{{\mathbf Q}_{2}}$: $004-q_{1z}$, $004+q_{1z}$, $006-q_{1z}$, and $006+q_{1z}$ for ${\mathbf Q}_{1}$, and $004$ and $006$ for ${\mathbf Q}_{2}$.
Note that, $f({\mathbf Q}_{\eta}) \equiv A \exp(-as^{2})+B \exp(-bs^{2})+C \exp(-cs^{2})+D$, where $A = 0.0755$, $a = 25.2960$, $B = 0.3001$, $b = 11.5993$, $C = 0.6438$, $c = 4.0252$, and $D = -0.0196$ for the Eu$^{2+}$ ion, and $s=1/2d_{hkl}$.
Using the lattice constant $c = 17.837~\AA$ at 10~K (Table~\ref{TabS1}) and the relation shown in Figs.~5(b) in the main text, we finally obtain $m_{{\mathbf Q}_{1}}/m_{{\mathbf Q}_{2}} \approx 0.91$.

\clearpage
\subsection*{\label{Sec9} \large Note 9. Theoretical calculations based on an effective spin Hamiltonian}

\subsubsection{\normalsize Bilinear-biquadratic model in real space: model 1}

Donoway {\it et al.} \cite{2023_Don} recently demonstrated that a spin Hamiltonian, including long-range Heisenberg exchange and four-spin exchange interactions in real space on a one-dimensional chain, can realize a broken-helix state as the ground state at zero field.
We here investigate the ground state of the same model under an in-plane magnetic field.
The spin Hamiltonian is given by
\begin{equation}
\label{Eq1}
\hspace{-0.3cm}{\mathcal{H}}=\sum_{n}\sum_{i}J_{n}{\mathbf S}_{i}\cdot{\mathbf S}_{i+n}+\sum_{i}[K_{1212}({\mathbf S}_{i}\cdot{\mathbf S}_{i+1})^2+K_{1223}({\mathbf S}_{i}\cdot{\mathbf S}_{i+1})({\mathbf S}_{i+1}\cdot{\mathbf S}_{i+2})]-h\sum_{i}S_{i}^{x}.\tag{S1}
\end{equation}
Here, we introduce the Ruderman-Kittel-Kasuya-Yosida (RKKY) interaction $J_{n} = -J_{1}[\cos(k_{1}n)/(k_{1}n)^{d}] - J_{2}[\cos(k_{2}n)/(k_{2}n)^{d}]$ up to the 10th nearest neighbor ($n = 10$), where we set $J_{1} = 1$ as the energy unit, and $k_{1}=0.29\pi$ and $k_{2}=\pi$ in line with our RXS data.
Since the conditions $d = 2$ and $d = 3$ cannot stabilize an incommensurate magnetic modulation but instead result in a ferromagnetic state, we need to set $d=1$ \cite{2023_Don}.
The second and third terms in Eq.~(\ref{Eq1}) are the four-spin exchanges with $K_{1212}, K_{1223} > 0$, which arise from perturbative expansions of the spin-charge coupling in the Kondo lattice model.
The last term in Eq.~(\ref{Eq1}) represents the Zeeman coupling to an in-plane magnetic field $h \parallel x$.
We assume $XY$ spins with $|{\mathbf S}_{i}| = 1$.

\begin{figure}[h]
\centering
\renewcommand{\thefigure}{S\arabic{figure}}
\includegraphics[width=0.7\linewidth]{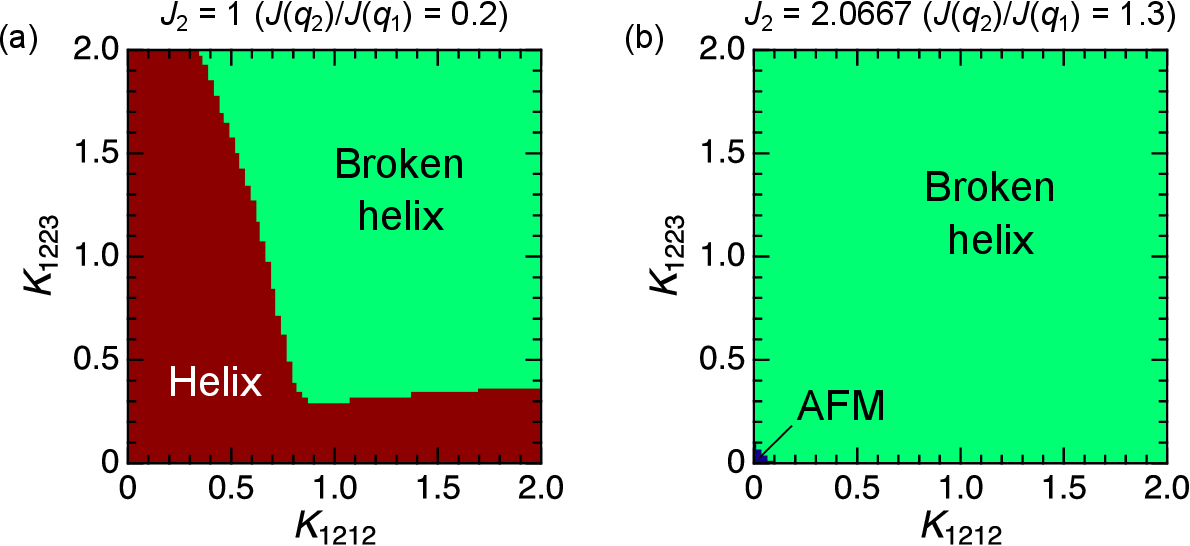}
\caption{Ground-state phase diagram of Eq.~(\ref{Eq1}) as a function of $K_{1212}$ and $K_{1223}$ at zero field for (a) $J_{2} = 1$ ($J(q_{2})/J(q_{1}) = 0.2$) and (b) $J_{2} = 2.0667$ ($J(q_{2})/J(q_{1}) = 1.3$). ``Helix'' represents a single-${\mathbf Q}$ helix state, which appears in a wide parameter ranges of $K_{1212}$ and $K_{1223}$ for $J_{2} = 1$.}
\label{FigS6}
\end{figure}

We performed simulated annealing with $N = 200$ spins for several parameter sets on Eq.~(\ref{Eq1}).
In this process, the parameter $J_{2}$ was tuned according to the structure of the exchange interactions in momentum space, $J(q)$.
Figures~\ref{FigS6}(a) and \ref{FigS6}(b) show zero-field ground-state phase diagrams as a function of $K_{1212}$ and $K_{1223}$ for $J_{2} = 1$ and $J_{2} = 2.0667$, respectively.
The ${\mathbf Q}_{1}$ modulation dominates in the former [Fig.~\ref{FigS6}(a)], where the ratio of $J(q)$ is $J(q_{2})/J(q_{1}) = 0.2$.
In this case, the introduction of finite $K_{1223}$ value is necessary to stabilize a double-${\mathbf Q}$ broken-helix state, as mentioned in Ref.~\cite{2023_Don}.
Conversely, the ${\mathbf Q}_{2}$ modulation is dominant in the latter [Fig.~\ref{FigS6}(b)], where $J(q_{2})/J(q_{1}) = 1.3$.
In this case, a double-${\mathbf Q}$ broken-helix state can be stabilized without introducing the $K_{1223}$ term.

\vspace{+0.5cm}

\begin{figure}[t]
\centering
\renewcommand{\thefigure}{S\arabic{figure}}
\includegraphics[width=0.8\linewidth]{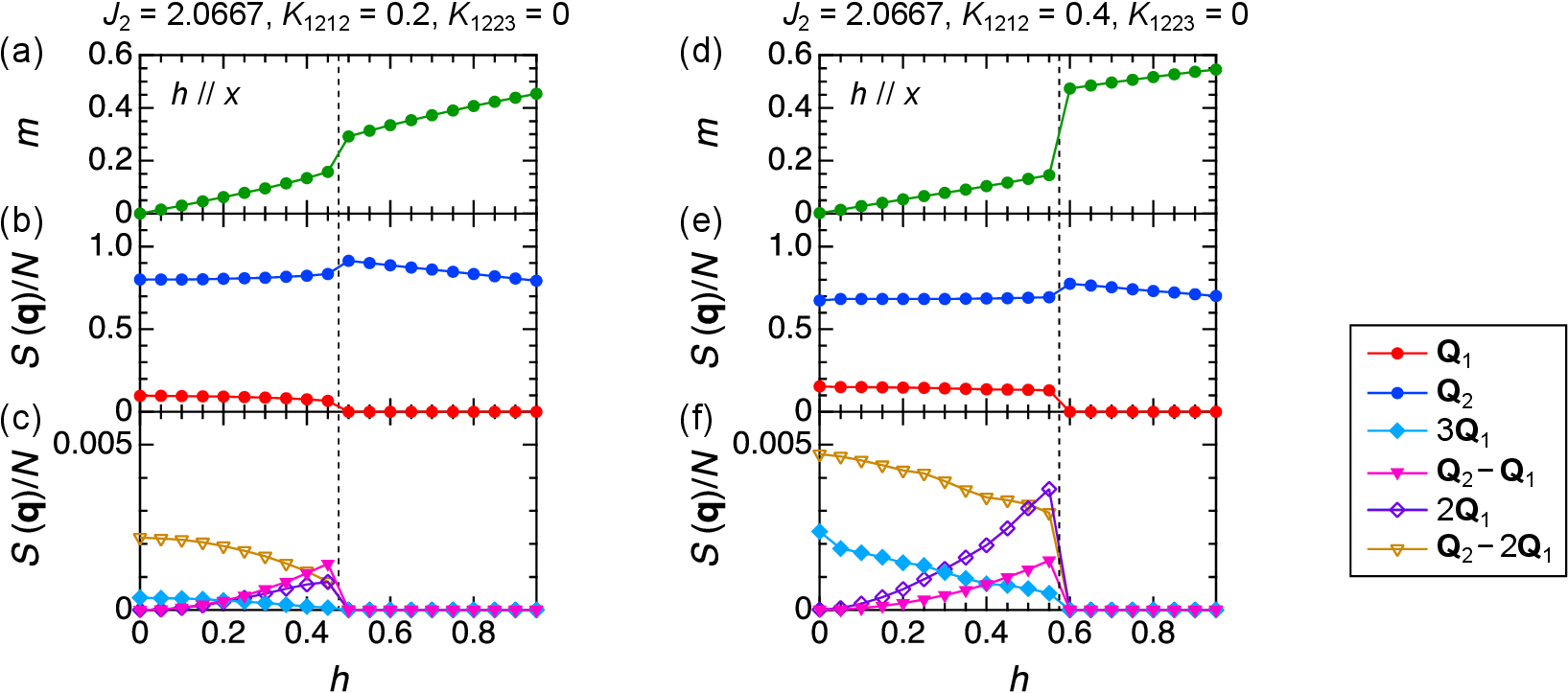}
\caption{Magnetic-field dependence of magnetization $m$ (top) and spin structure factor $S({\mathbf q})$ per a spin at several ${\mathbf Q}$ positions (bottom) for $h \parallel x$, calculated for Eq.~(\ref{Eq1}) with $J_{2} = 2.0667, K_{1212} = 0.2, K_{1223} = 0$ [(a)--(c)] and $J_{2} = 2.0667, K_{1212} = 0.4, K_{1223} = 0$ [(d)--(f)].}
\label{FigS7}
\vspace{+0.5 cm}
\end{figure}

\vspace{+0.5cm}
Figure~\ref{FigS7} shows the magnetic-field dependence of magnetization $m$ and spin structure factor, defined as $S({\mathbf q}) = (1/N)\sum_{i,j}\langle {\mathbf S}_{i} \cdot {\mathbf S}_{j}\rangle e^{i{\mathbf q} \cdot ({\mathbf r}_{i}-{\mathbf r}_{j})}$, at various ${\mathbf Q}$ positions for $J_{2} = 2.0667, K_{1212} = 0.2$, $K_{1223} = 0$ [(a)--(c)] and for $J_{2} = 2.0667, K_{1212} = 0.4$, $K_{1223} = 0$ [(d)--(f)].
We note that the moderate introduction of the $K_{1223}$ term does not significantly alter the results.
At $h = 0$, $S({\mathbf Q}_{1})/S({\mathbf Q}_{2}) \approx 0.13$ and 0.22 for $K_{1212} = 0.2$ and 0.4, respectively.
These values closely match our experimental observation of $I_{{\mathbf Q}_{1}}/I_{{\mathbf Q}_{2}} \approx 0.17$ for sample \#1 ($q_{1z} = 0.290$).
However, the ${\mathbf Q}_{1}$ component suddenly disappears in a high-field phase above a metamagnetic transition.
Furthermore, there is no discernible tendency for the enhancement of higher-harmonic modulations.
These results are in qualitative contradiction with our experimental observation of a double-${\mathbf Q}$ fanlike state, as shown in Figs.~3 and 4 in the main text.
We hence conclude that the magnetism in EuIn$_{2}$As$_{2}$ cannot be perfectly reproduced by the spin Hamiltonian Eq.~(\ref{Eq1}).

Here, we note that the condition $d = 1$ introduced in the Heisenberg term is not compatible with the three-dimensional character of the hole pocket observed in ARPES \cite{2020_Sat}.
It is likely that incorporating the Heisenberg exchange interactions only along the 1D chain is insufficient to adequately describe the RKKY interaction in real 3D systems. 

\clearpage

\subsubsection{\normalsize Bilinear-biquadratic model in real space: model 2}

As an alternative to Eq.~(\ref{Eq1}), we consider a more general one-dimensional spin model in a real space as follows: 
\begin{equation}
\label{Eq2}
\hspace{-0.3cm}{\mathcal{H}}=\sum_{n}\sum_{i}J_{n}{\mathbf S}_{i}\cdot{\mathbf S}_{i+n}+K_{1212}\sum_{i}({\mathbf S}_{i}\cdot{\mathbf S}_{i+1})^2-h\sum_{i}S_{i}^{x},\tag{S2}
\end{equation}
Here, $J_{n}$ can be considered as an effective exchange interaction between the $n$-th nearest-neighbor Eu layers, where all the RKKY interaction and the dipole-dipole interaction are renormalized.
Accordingly, in contrast to the case of Eq.~(\ref{Eq1}), we do not impose any restriction on $J_{n}$.
Additionally, the biquadratic interaction $K_{1212}$ is introduced to stabilize a double-${\mathbf Q}$ state.

\begin{figure}[h]
\centering
\renewcommand{\thefigure}{S\arabic{figure}}
\vspace{+0.5 cm}
\includegraphics[width=0.8\linewidth]{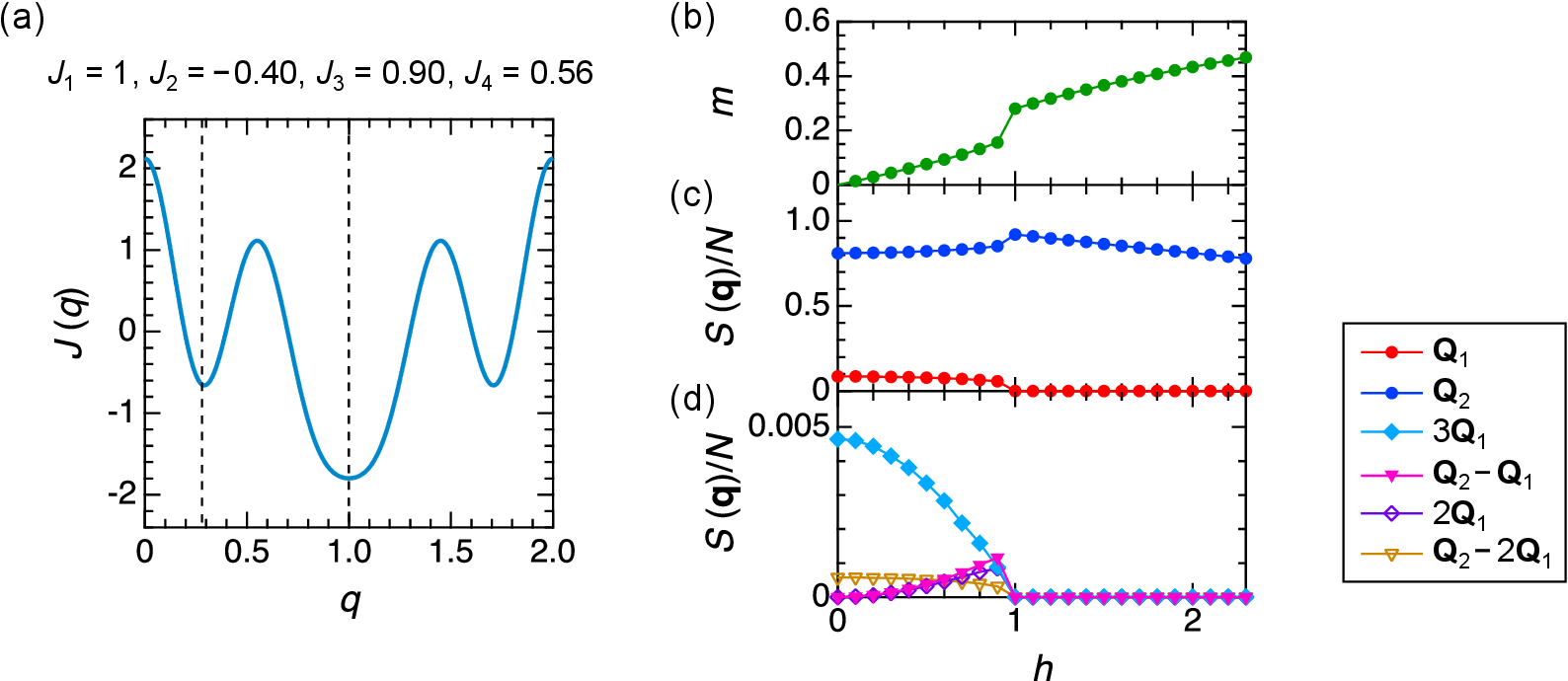}
\caption{(a) Energy landscape of $J(q)$ as a function of $q$. [(b)--(d)] Magnetic-field dependence of (b) magnetization $m$ and (c)(d) spin structure factor $S({\mathbf q})$ per a spin at several ${\mathbf Q}$'s for $h \parallel x$ in Eq.~(\ref{Eq2}). All the data are obtained with $J_{2} = -0.40$, $J_{3} = 0.90$, $J_{4} = 0.56$, $K_{1212} = 0.6$.}
\label{FigS8}
\end{figure}

We reveal through an exhaustive parameter search that $n \geq 4$ is necessary for stabilizing a double-${\mathbf Q}$ broken-helix state with $q_{1} = 0.29$ and $q_{2} = 1$ at zero field.
One typical parameter set is $J_{1} = 1$, $J_{2} = -0.40$, $J_{3} = 0.90$, $J_{4} = 0.56$, $K_{1212} = 0.6$, as shown in Fig.~\ref{FigS8}(a).
Nevertheless, the ${\mathbf Q}_{1}$ component disappears after a metamagnetic transition [Figs.~\ref{FigS8}(b)--\ref{FigS8}(d)], as in the case of Eq.~(\ref{Eq1}).
Hence, we conclude that the magnetic phases in EuIn$_{2}$As$_{2}$ cannot be adequately described by a simple spin Hamiltonian based on the real space picture.

\clearpage
\subsubsection{\normalsize Bilinear-biquadratic model in momentum space}

\begin{figure}[h]
\centering
\renewcommand{\thefigure}{S\arabic{figure}}
\includegraphics[width=0.75\linewidth]{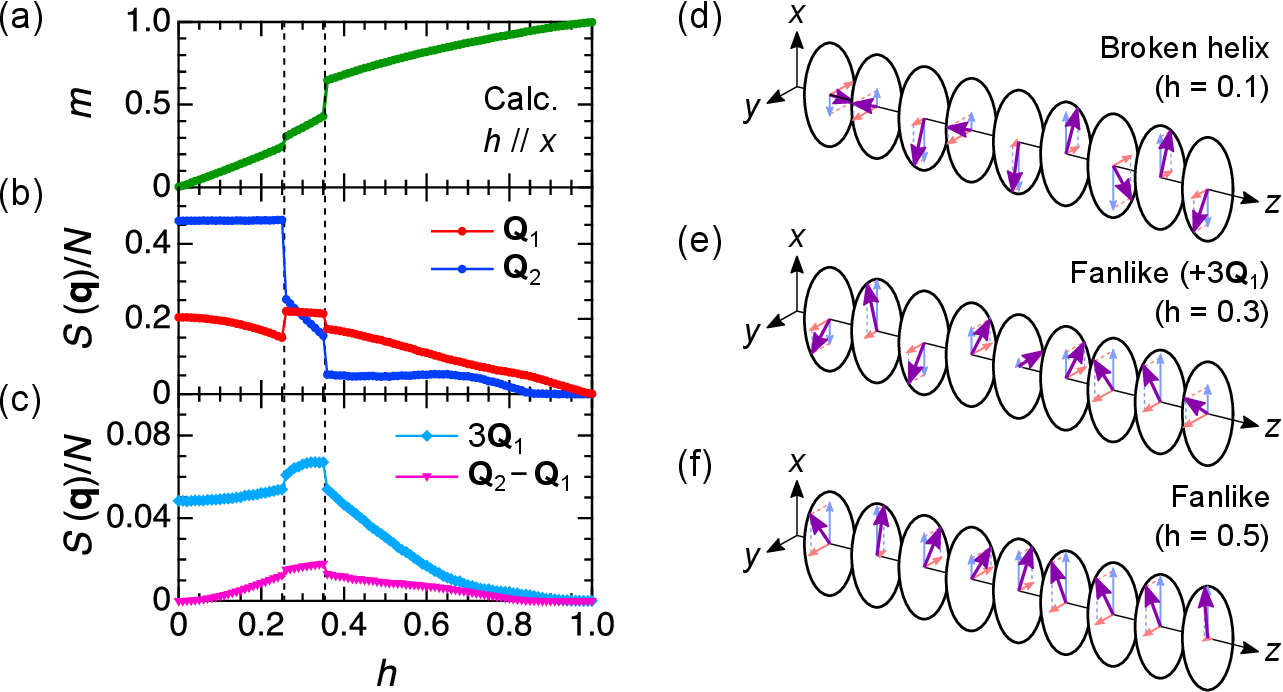}
\caption{Magnetic-field dependence of (a) magnetization $m$ and (b)(c) spin structure factor $S({\mathbf q})$ per a spin at several ${\mathbf Q}$'s for $h \parallel x$, calculated for Eq.~(\ref{Eq3}) with $K = 0.9$, $\alpha = 0.48$, and $\alpha' = 0.9$. Local spin configurations at $h = 0.1$, 0.3, and 0.5 are schematically illustrated by purple arrows in panels (d), (e), and (f), respectively. The light blue (red) arrow indicates the $x$ ($y$) component of each spin.}
\vspace{-0.2cm}
\label{FigS9}
\end{figure}

Here, we employ an effective spin Hamiltonian composed of exchange interactions in momentum space:
\begin{equation}
\begin{split}
\label{Eq3}
\hspace{-0.3cm}{\mathcal{H}}=&-J{\mathbf S}_{{\mathbf Q}_{1}}\cdot{\mathbf S}_{-{\mathbf Q}_{1}}-J'{\mathbf S}_{{\mathbf Q}_{2}}\cdot{\mathbf S}_{-{\mathbf Q}_{2}}-J''{\mathbf S}_{{\mathbf Q}_{3}}\cdot{\mathbf S}_{-{\mathbf Q}_{3}}\\&+K({\mathbf S}_{{\mathbf Q}_{1}}\cdot{\mathbf S}_{-{\mathbf Q}_{1}})^{2}+K'({\mathbf S}_{{\mathbf Q}_{2}}\cdot{\mathbf S}_{-{\mathbf Q}_{2}})^{2}+K''({\mathbf S}_{{\mathbf Q}_{3}}\cdot{\mathbf S}_{-{\mathbf Q}_{3}})^{2}\\&-K_{2}({\mathbf S}_{{\mathbf Q}_{1}}\cdot{\mathbf S}_{{\mathbf Q}_{2}})({\mathbf S}_{-{\mathbf Q}_{1}}\cdot{\mathbf S}_{-{\mathbf Q}_{2}})-h\sum_{i}S_{i}^{x},
\raisetag{0.77cm}
\end{split}
\tag{S3}
\end{equation}
where we take into account dominant interactions with ${\mathbf Q}_{1} = 0.29\pi$ and ${\mathbf Q}_{2} = \pi$.
In addition to the conventional bilinear and biquadratic terms \cite{2017_Hay}, Eq.~(\ref{Eq3}) includes an intertwined coupling term ($K_{2}$) with a negative sign, which plays a crucial role in stabilizing a double-${\mathbf Q}$ fanlike state in high magnetic fields.
We also introduce higher-harmonic exchange coupling terms ($J''$ and $K''$) with ${\mathbf Q}_{3} = 3{\mathbf Q}_{1}$ to reproduce the enhancement of the $3{\mathbf Q}_{1}$ modulation immediately after the metamagnetic transition.
The last term in Eq.~(\ref{Eq3}) represents the Zeeman coupling to an in-plane magnetic field $h \parallel x$.
For simplicity, we set $J'=\alpha J$, $J'' = \alpha' J$, $K' = \alpha^{2} K$, $K'' = \alpha'^{2} K$, $K_{2} = K/2$, $J = 1$ as the energy unit, and consider $XY$ spins with $|{\mathbf S}_{i}| = 1$.
We performed simulated annealing with $N = 200$ spins for several parameter sets, and calculated the magnetization $m$ and spin structure factor $S({\mathbf q}) = (1/N)\sum_{i,j}\langle {\mathbf S}_{i} \cdot {\mathbf S}_{j}\rangle e^{i{\mathbf q} \cdot ({\mathbf r}_{i}-{\mathbf r}_{j})}$.
Figures~\ref{FigS9}(a)--\ref{FigS9}(c) show one typical result, which qualitatively reproduces our RXS data (see Fig.~3 in the main text).
The system undergoes metamagnetic transitions, accompanied by the enhancement of $S(3{\mathbf Q}_{1})$ and $S({\mathbf Q}_{2} - {\mathbf Q}_{1})$.
We find that other parameter sets bring about qualitatively distinct behaviors in higher harmonics, such as the enhancement of $S({{\mathbf Q}_{2} - 2{\mathbf Q}_{1}})$ instead of $S(3{\mathbf Q}_{1})$.

Figures~\ref{FigS9}(d)-\ref{FigS9}(f) show schematics of local spin configurations in three different phases.
The spin configuration at $h = 0.3$ is far from the conventional fanlike state in that some spins still possess a negative $x$ component.
Notably, a square-wave-shaped modulation appears in the $y$ component (perpendicular to $h$), reflecting the effect of $3{\mathbf Q}_{1}$ along with ${\mathbf Q}_{1}$.
At $h = 0.5$, the spin configuration is like a conventional fanlike state, while the remaining ${\mathbf Q}_{2}$ component contributes to a complex spin-flipping pattern.
Although our magnetization data [Fig.~3(a) in the main text] shows no clear signs of the second transition above $H_{\rm c1}$, the magnetization data reported in Fig.~3a in Ref.~\cite{2023_Soh} clearly exhibits a second anomaly in $dM/dB$ at around 0.3~T.
Therefore, our theoretical picture would qualitatively describe the field evolution of a magnetic structure in EuIn$_{2}$As$_{2}$.

\vspace{+1.0cm}
